\pdfoutput=1
\documentclass[sigconf,screen]{acmart}

\def\FIGDIR{./figures}          

\newcommand{\forsubmission}{}

\input{include/packages}

\newcommand{\insertFigure}[2]{
    \begin{figure}[t]
        \centering
        \includegraphics[width=\linewidth]{\FIGDIR/#1.pdf}
        \caption{#2}
        \label{fig:#1}
    \end{figure}
}

\newcommand{\insertWideFigure}[2]{
    \begin{figure*}[h]
        \centering
        \includegraphics[width=\textwidth]{\FIGDIR/#1.pdf}
        \caption{#2}
        \label{fig:#1}
    \end{figure*}
}

\ifx\forsubmission\undefined
\newcommand{\TODO}[1]{\textcolor{red}{TODO: #1}}
\newcommand{\SK}[1]{\textcolor{red}{SK: #1}}
\newcommand{\HK}[1]{\textcolor{blue}{HK: #1}}

\newcommand{\revision}[1]{\textcolor{blue}{#1}}
\newcommand{\fixme}[1]{{\color{red} {#1}}}
\newcommand{\tobechecked}[1]{\textcolor{blue}{#1}}
\else
\newcommand{\TODO}[1]{\textcolor{red}{}}
\newcommand{\SK}[1]{\textcolor{red}{}}
\newcommand{\HK}[1]{\textcolor{blue}{}}
\newcommand{\revision}[1]{#1}
\newcommand{\fixme}[1]{{\color{red} {}}} 
\newcommand{\tobechecked}[1]{#1}
\fi

\newcommand{\squishlist}{
 \begin{list}{$\bullet$}
  { \setlength{\itemsep}{0pt}
     \setlength{\parsep}{3pt}
     \setlength{\topsep}{3pt}
     \setlength{\partopsep}{0pt}
     \setlength{\leftmargin}{1.5em}
     \setlength{\labelwidth}{1em}
     \setlength{\labelsep}{0.5em} } }

\newcommand{\squishlisttwo}{
 \begin{list}{$\bullet$}
  { \setlength{\itemsep}{0pt}
     \setlength{\parsep}{0pt}
    \setlength{\topsep}{0pt}
    \setlength{\partopsep}{0pt}
    \setlength{\leftmargin}{2em}
    \setlength{\labelwidth}{1.5em}
    \setlength{\labelsep}{0.5em} } }

\newcommand{\squishend}{
  \end{list}  }

\newcommand{\betterparagraph}[1]{\noindent \textbf{#1. }}

\newcommand{\scheduler}{\textsc{DREAM}\xspace}

\newcommand{\scoreMetric}{\textsc{MapScore}\xspace}

\newcommand{\costFunction}{\textsc{UXCost}\xspace}

\newcommand{\vr}{\texttt{VR\_Gaming}\xspace}

\newcommand{\ar}{\texttt{AR\_Social}\xspace}

\newcommand{\dyscore}{\textsc{DREAM-MapScore}\xspace}
\newcommand{\smartdrop}{\textsc{DREAM-SmartDrop}\xspace}
\newcommand{\full}{\textsc{DREAM-Full}\xspace}

\newcommand{\cmark}{\ding{51}}%
\usepackage{hhline}
\usepackage{amsmath, amsfonts}
\usepackage{graphicx}
\usepackage{textcomp}
\usepackage{xcolor}
\usepackage{float}
\usepackage{url}
\usepackage{algorithm}
\usepackage[noend]{algpseudocode}
\usepackage{makecell}
\usepackage{soul}
\usepackage{hyperref}
\usepackage{xparse}
\usepackage{natbib}
\usepackage{balance}
\usepackage{xspace}
\usepackage{framed, multido}

\AtBeginDocument{%
  \providecommand\BibTeX{{%
    \normalfont B\kern-0.5em{\scshape i\kern-0.25em b}\kern-0.8em\TeX}}}

\makeatletter
\NewDocumentCommand{\LeftComment}{s m}{%
  \Statex \IfBooleanF{#1}{\hspace*{\ALG@thistlm}}\(\triangleright\) #2}
\makeatother
\algnewcommand{\LeftCommentFirst}[2]{\Statex\hspace{#1} \(\triangleright\) #2}

\RequirePackage[T1]{fontenc} \RequirePackage[varqu]{zi4} \RequirePackage[libertine]{newtxmath}
\algdef{SE}[SUBALG]{Indent}{EndIndent}{}{\algorithmicend\ }%
\algtext*{Indent}
\algtext*{EndIndent}

\copyrightyear{2024}
\acmYear{2024}
\setcopyright{acmcopyright}
\acmConference[ASPLOS '24, April 27-May 1]{Architectural Support for Programming Languages and Operating Systems}{2024}{San Diego, CA, USA}
\acmBooktitle{The 29th Conference on Architectural Support for Programming Languages and Operating Systems (ASPLOS'24)}
\acmDOI{10.1145/3623278.3624753}

\begin{document}

\setlength{\abovedisplayskip}{2pt}
\setlength{\belowdisplayskip}{5pt}
\setlength{\abovedisplayshortskip}{0pt}
\setlength{\belowdisplayshortskip}{0pt}

\title{DREAM: A Dynamic Scheduler for \underline{D}ynamic \underline{Rea}l-time \underline{M}ulti-model ML Workloads}

\author{Seah Kim}
\authornote{Work done during Summer Internship at Meta}
\affiliation{%
  \institution{UC Berkeley}
  \city{Berkeley}
  \state{CA}
  \country{USA}
}
\email{seah@berkeley.edu}
\author{Hyoukjun Kwon}
\authornote{Corresponding author}
\affiliation{%
  \institution{UC Irvine, Meta}
  \city{Irvine}
  \state{CA}
  \country{USA}
}
\email{hyoukjun.kwon@uci.edu}
\author{Jinook Song}
\affiliation{%
  \institution{Meta}
  \city{Sunnyvale}
  \state{CA}
  \country{USA}
}
\email{jinooksong@meta.com}
\author{Jihyuck Jo}
\affiliation{%
  \institution{Meta}
  \city{Sunnyvale}
  \state{CA}
  \country{USA}
}
\email{jjo@meta.com}
\author{Yu-Hsin Chen}
\affiliation{%
  \institution{Meta}
  \city{Sunnyvale}
  \state{CA}
  \country{USA}
}
\email{yhchen@meta.com}
\author{Liangzhen Lai}
\affiliation{%
  \institution{Meta}
  \city{Sunnyvale}
  \state{CA}
  \country{USA}
}
\email{liangzhen@meta.com}
\author{Vikas Chandra}
\affiliation{%
  \institution{Meta}
  \city{Sunnyvale}
  \state{CA}
  \country{USA}
}
\email{vchandra@meta.com}


\begin{abstract}
Emerging real-time multi-model ML (RTMM) workloads such as AR/VR and drone control involve dynamic behaviors in various granularity; task, model, and layers within a model.
Such dynamic behaviors introduce new challenges to the system software in an ML system since the overall system load is not completely predictable, unlike traditional ML workloads. 
In addition, RTMM workloads require real-time processing, involve highly heterogeneous models, and target resource-constrained devices.
Under such circumstances, developing an effective scheduler gains more importance to better utilize underlying hardware considering the unique characteristics of RTMM workloads.
Therefore, we propose a new scheduler, \scheduler, which effectively handles various dynamicity in RTMM workloads targeting multi-accelerator systems.
\scheduler quantifies the unique requirements for RTMM workloads and utilizes the quantified scores to drive scheduling decisions, considering the current system load and other inference jobs on different models and input frames.
\scheduler utilizes tunable parameters that provide fast and effective adaptivity to dynamic workload changes.
In our evaluation of five scenarios of RTMM workload, \scheduler reduces the overall \costFunction, which is an equivalent metric of the energy-delay product (EDP) for RTMM defined in the paper, by 32.2\% and 50.0\% in the geometric mean (up to 80.8\% and 97.6\%) compared to state-of-the-art baselines, which shows the efficacy of our scheduling methodology.

\end{abstract}


\maketitle

\section{Introduction}
\label{sec:introduction}

\insertWideFigure{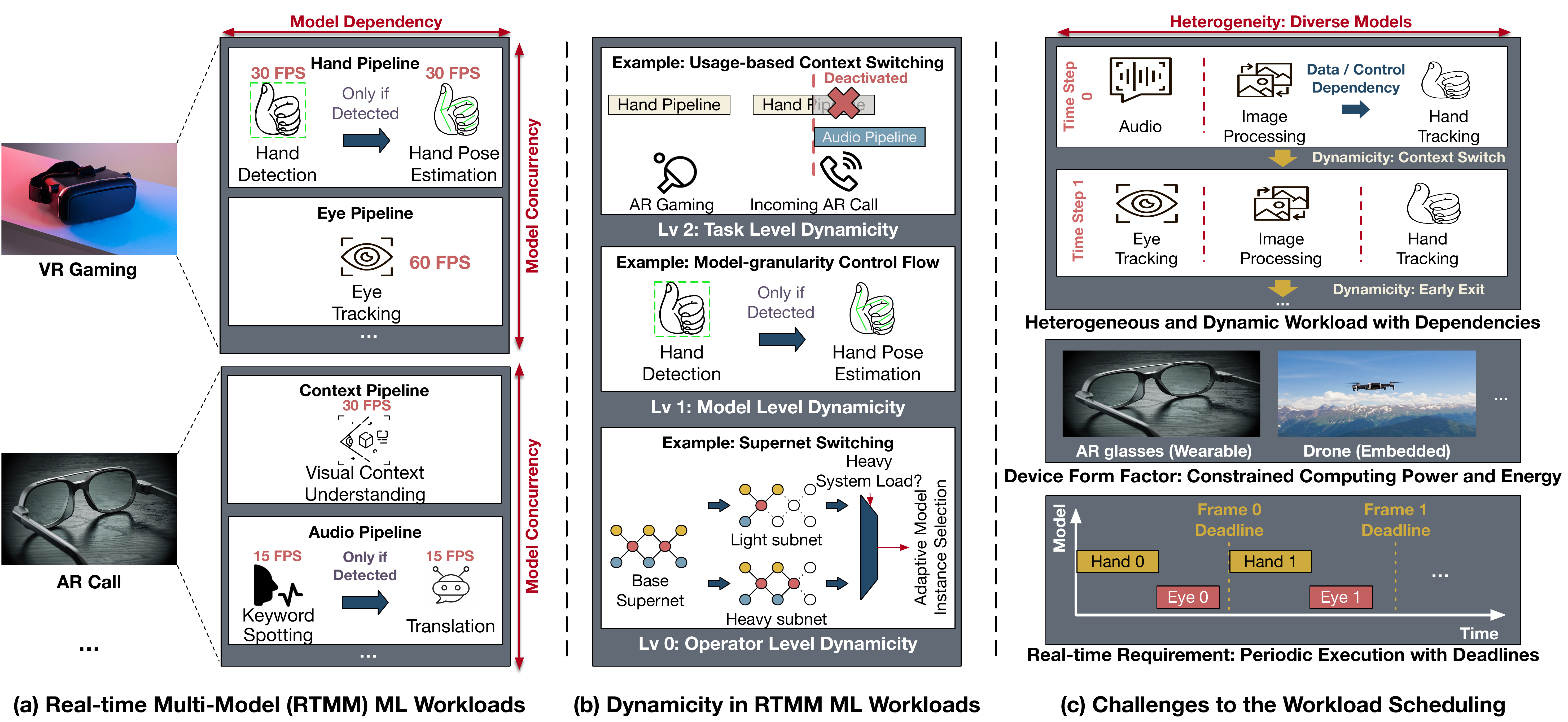}{A summary of the motivation for this work. (a) Example real-time multi-model ML workloads that have multiple concurrent pipelines and cascaded models within some pipelines, which adds control and data dependencies to the scheduling consideration. (b) Three levels of dynamicity found in real-time multi-model workloads and examples of each. (c) Challenges to the scheduler from workloads and dynamicity. \vspace{-4mm}}

As ML-based applications become more diverse, ML inference workloads in emerging applications such as augmented and virtual reality (AR/VR) deploy many ML models with complex dependency and concurrency~\cite{xrbench}, as shown in ~\autoref{fig:Motivation} (a).
Such applications often require real-time processing, which lead to real-time multi-model (RTMM) ML inference workloads (e.g., drone navigation~\cite{trailnet}).
RTMM workloads impose unique requirements that distinguish them from previous ML inference workloads often based on one or several independent models without model level dependency and concurrency~\cite{MLperf}.
As summarized in~\autoref{fig:Motivation} (c), such challenges include (1) highly heterogeneous ML models (e.g., model size, operators, and tensor size) from diverse tasks and multi-modal sensor inputs, (2) rich dynamicity in various levels as illustrated in~\autoref{fig:Motivation} (b), (3) complex model level data and control dependencies, (4) constrained computing power and energy in target devices (e.g., AR glasses), and (5) real-time requirements due to continuous and periodic processing of tasks with deadlines.


Among these challenges, dynamic workloads can easily lead to unpredictable system loads, imposing a new challenge for ML systems that previously relied on highly deterministic latency information for scheduling decisions~\cite{Planaria, kwon2021heterogeneous}. 
In addition, the dynamicity exists in various levels of granularity; across task (i.e., which models to include in the workload), model (i.e., which version of a model to run~\cite{alphanet, once-for-all}), and operator (i.e., which layers to run within a model) levels.
Combined with the real-time requirements on constrained hardware resources, diverse dynamicity becomes a non-trivial challenge for real-time ML systems, in particular for schedulers.
Although some static scheduling methods have been proposed~\cite{kwon2021heterogeneous} for multi-model workloads, they are not tailored for the RTMM workloads.
Prior works have proposed dynamic scheduling-based methods, which shed light on a part of the challenges targeting multi-tenancy~\cite{Planaria, Veltair, moca, PREMA} within a task.
However, they are not tailored for the dynamicity of RTMM workloads and other unique challenges of RTMM workloads, as summarized in~\autoref{tab:rtmm-breakdown}.

To address new challenges from RTMM, we propose \scheduler (Scheduler for \textbf{D}ynamic and \textbf{Rea}l-time \textbf{M}ulti-model Multi-Task ML Workloads), which targets ML accelerator-based systems and holistically considers all the main challenges of emerging RTMM workloads: real-time requirements (FPS, deadlines), concurrent processing of multiple tasks with cascaded models, and adapting to dynamic workload changes.
For the real-time processing and concurrency challenge, we propose a score metric named \scoreMetric that considers both urgency and fairness, which facilitates optimization not only for task-specific performance but also for overall performance across all tasks. 
For the complex dependency challenge of cascaded models, \scheduler tracks the model dependency within an input frame and across multiple frames.
For the dynamicity challenge, we develop a dynamic scheduling method with tunable parameters that provides fast and effective adaptivity to workload changes.
\scheduler also supports a variety of ML systems based on accelerators, even ones including multiple accelerators with heterogeneous size and dataflow like heterogeneous dataflow accelerators proposed in Herald~\cite{kwon2021heterogeneous}.

We evaluate \scheduler using \costFunction introduced in~\autoref{subsec:mapscore}, which attempts to quantify the overall user experience using deadline violation rate and energy consumption.
On average, \scheduler reduces \costFunction by 32.2\% (up to 80.8\%) and 50.0\% (up to 97.6\%) compared to state-of-the-art baselines Planaria~\cite{Planaria} and Veltair~\cite{Veltair}, respectively, on industry-originated RTMM workloads~\cite{xrbench,trailnet}.

We summarize our contributions as follows:

\squishlist
{\item A score tailored for driving scheduling decisions, which thoroughly captures key requirements and unique characteristics of RTMM workloads.}
{\item A dynamic scheduler with tunable parameters and an online tuning method that provides fast adaptivity for workload changes without blocking the execution of workloads.}
{\item A preemptive frame drop method that proactively drops a frame early when a deadline violation is expected, which facilitates global optimization across frames and models.}
{\item Exploration of Supernet switching~\cite{once-for-all} in the context of RTMM that leverages a weight-sharing Supernet to improve ML system schedulers by dynamically switching to lighter model variants under heavy system loads, which also facilitates the optimization in a global scope.}
{\item Case studies on both homogeneous and heterogeneous hardware accelerator systems with different sizes and dataflows running industry-originated realistic RTMM ML workloads, which provide new insights on the scheduling problem for RTMM workloads.}


\squishend

\section{Background and Motivation}
\label{sec:background}

Multi-task multi-model ML workloads have emerged from complex applications utilizing ML for various sub-tasks.
Among them, applications such as AR/VR and autonomous driving require real-time processing, which constructed a new class of ML workload, real-time multi-model (RTMM) ML workloads~\cite{xrbench}.
We discuss their unique characteristics and implications for the ML system design.

\subsection{Characteristics of RTMM Workloads}
\label{sec:RTMM_workloads}

Traditional ML workloads run inferences on a single model for a single user or a collection of different single-model workloads for many users (i.e., multi-tenancy).
Unlike such workloads, RTMM workloads require to (1) run multiple models in a cascaded manner with inter-model dependencies (e.g., hand pipeline in~\autoref{fig:Motivation} (a)), (2) concurrently run multiple ML pipelines, (3) meet real-time requirements (i.e., meeting deadlines for periodic inferences), and (4) support dynamic workloads that change based on user inputs or user context changes. We discuss each challenge in detail as follows.

\betterparagraph{Cascaded models (ML pipeline)}
ML pipelines, which cascade multiple models to perform more sophisticated tasks (e.g., eye tracking pipeline: cascaded eye segmentation and gaze estimation models~\cite{you2022eyecod}), have emerged to solve complex ML problems.
Such ML pipelines with cascaded models add dependency across multiple models, which is one of the key differences from traditional multi-tenancy ML workloads.

\betterparagraph{Concurrent ML pipelines}
Complex applications such as AR are based on diverse tasks. 
For example, a VR game can require both hand- and eye-tracking pipelines concurrently to provide a highly immersive and interactive user experience~\cite{xrbench}.
That is, hand- and eye-tracking pipelines need to be executed concurrently for such an application, which introduces the concurrency challenge.

\betterparagraph{Real-time Requirement}
Applications that involve active interaction with a user (e.g., VR) or environment (e.g., drone control) require real-time processing of ML workloads.
The real-time requirement can be translated into three core system requirements: periodically streamed input data, target processing rate (FPS), and processing deadline for each frame.

\betterparagraph{Dynamicity}
Another notable characteristic of emerging RTMM workloads is the dynamicity: the workload changes based on the user inputs and environment.
For example, if a drone flying in a building moves out from the building, the navigation ML model should be updated from an indoor environment-oriented to a model optimized for outdoor environments.
Such changes can occur at diverse granularity, including at task (when the task changes, the model changes accordingly), model (model cascade pipeline with dependency), and operator (Supernet model variant, branchy early-exit behavior) levels, as illustrated in~\autoref{fig:Motivation} (b).
As we identify the dynamicity as one of the core challenges in RTMM workloads, we discuss the dynamicity in detail next.

\subsection{Dynamicity in Workloads}
\label{sec:dynamicity}


We introduce three levels of dynamicity found in RTMM workloads and discuss why the static scheduling approach would fail due to such dynamicity.
The sources of dynamicity in the ML workloads range from the intra-model to the task levels within an RTMM workload, as illustrated in~\autoref{fig:Motivation} (b).

\betterparagraph{Task Level Dynamicity} As the RTMM workload is a real-time workload, the user context and usage scenario can change over time.
For instance, a user playing a hand-interaction-based VR game would completely change the usage scenario to an AR call when the user receives an incoming AR call.
Such a usage scenario change triggers a context switch to a totally different set of ML pipelines and models.
In such a case, the new pipeline has to be scheduled immediately, and the rest of the models in the pipeline need to be flushed or dropped.
This is a major challenge to static algorithm-based schedulers because they need to stop the execution and re-construct an entire schedule every time a new set of workloads is identified.

\betterparagraph{Model Level Dynamicity}
\label{sec:model_dynamicity}
In ML pipelines that include cascaded models, the execution of a model in an ML pipeline depends on the results of prior models of the ML pipeline.
\tobechecked{When the dependency is control dependency, the workload becomes non-deterministic as the results are only available after running the prior model and vary based on inputs to the ML pipeline.
For such non-deterministic workloads, it is infeasible for static schedulers to generate a valid and optimized schedule in advance as the actual workload is unknown at the scheduling time.}


\betterparagraph{Operator Level Dynamicity}
Supernet~\cite{alphanet} refers to an emerging class of models that have large base model structures where their subsets (i.e., sub-network) are selected for different deployment scenarios.
Supernet facilitates the training of multiple models with a single training process, which provides scalability for the model development process.
Recent works such as Once-for-all~\cite{once-for-all} utilize the Supernet approach to train multiple versions of a model in the model size - model performance (e.g., accuracy) trade-off space.
For the ML system, such an approach allows an optimization technique to select the best version of a Supernet-based model depending on the system status (e.g., overall system load, thermal, etc.) and application requirements (e.g., face recognition for unlocking a phone requires very high accuracy). 
Although such an emerging optimization technique is an effective approach, it introduces another dynamicity within a model, which increases the complexity of the scheduling problem.

Unlike previous work that selects the model instance offline~\cite{tong2022enabling}, we explore a new way to utilize Supernet-based models~\cite{once-for-all} for better scheduling decisions in the RTMM context, which we discuss in detail in~\autoref{subsubsec:supernet_switching}.
Beyond Supernet-based models, dynamicity can originate from models utilizing control flow-based techniques to select the best path based on intermediate scores. Examples include early-exit~\cite{branchynet, RAPID-RL} and layer-skipping~\cite{blockdrop, SkipNet, conv-aig} methods.
The goal of those models is to select the optimal computation-graph traversal in the accuracy-compute overhead trade-off space. Due to their dynamic nature, static scheduling would not be able to leverage such techniques, leading to conservative scheduling targeting worst cases (i.e., the longest path) to ensure correctness.



\subsection{Limitation of Static Scheduling} 
\label{sec:motivation}

\begin{figure}[t]
    \centering
    \includegraphics[width=0.85\linewidth]{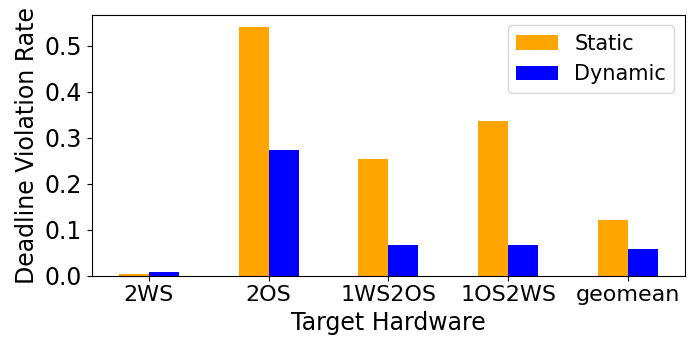}
    \vspace{-3mm}
    \caption{The deadline violation rate on \texttt{AR\_Call} workload used in the evaluation (\autoref{tab:workload_scenarios}) using static and dynamic first-come-first-served (FCFS) schedulers.\vspace{-2mm}}
    \label{fig:motiv_fail_rate}
\end{figure}

\autoref{fig:motiv_fail_rate} shows the deadline violation rate of static and dynamic first-come-first-served (FCFS) schedulers using the \texttt{AR\_Call} workload in four accelerator styles listed in~\autoref{tab:target_hw_settings}.
We select the \texttt{AR\_Call} scenario as it has both an audio pipeline and a dynamic model, SkipNet~\cite{SkipNet}.
We assume a 50\% probability for the positive keyword spotting results in the audio pipeline and a 50\% probability of skipping layers for SkipNet, which provides 72\% of Top-1 accuracy on ImageNet~\cite{SkipNet}.

As \autoref{fig:motiv_fail_rate} shows, even if we apply the same simple FCFS algorithm, dynamic scheduling decreases the deadline violation rate by 52.9\%, on average.
This presents a good motivation for designing a dynamic scheduler for RTMM workloads.
Although we focused on the deadline violation rates considering the real-time requirements, there are many other factors to consider to design an effective dynamic scheduler, which includes energy, hardware heterogeneity, and usage scenario-scope optimization (not local optimization for each model).
To holistically consider all the aspects, we define a set of scoring metrics and utilize the metric to drive scheduling decisions. 
We discuss the details of our scoring metric next.

\section{Scoring Metrics}
\label{sec:score}
In this section, we introduce our scoring metric that considers (1) \textbf{Urgency}, the time margin to a deadline modeling real-time requirements, (2) \textbf{Preference}, hardware heterogeneity and their preference to different ML operators (or layers) modeling the heterogeneity, (3) \textbf{Starvation}, the degree of starvation of each model, and (4) \textbf{Energy}, estimated energy consumption for running an operator that considers the constrained energy in target devices of RTMM workloads (e.g., AR glasses). 
Based on the unit score metrics, we introduce \scoreMetric as a comprehensive metric that captures all the important requirements considered in the four unit scores. 

\begin{algorithm}[t]
\caption{\scoreMetric computation for an inference task} \label{alg:mapscore}
\begin{algorithmic}[1]

\LeftComment{\textbf{Inputs}> \textbf{$\bm{acc}$}(accelerator ID), \textbf{$\bm{tsk}$} (inference task ID), \textbf{\textit{AccList}} (A list of accelerators), $\bm{N_{acc}}$ (The number of accelerators), $\bm{T_{curr}}$ (current time), $\bm{T_{cmpl}}$ (A list of the lastly scheduled layer completion time for each task), $\bm{Q_{task}}$ (Queues for remaining layers for each task), $\bm{T_{deadline}}$ (A list of deadlines for each task), $\bm{Est_{Latency}}$ and $\bm{Est_{Energy}}$ (Estimated latency and energy for each accelerator and layer pair generated offline by a cost model or a simulator.), $\bm{Stack_{task}}$ (stacks for each task tracking the completed layers and accelerators executed each layer)}
\LeftComment{\textbf{Output}> $\mathrm{MapScore}(tsk, acc)$: \scoreMetric for $tsk$ on accelerator $acc$}
    \State $\bm{Function}\ \mathrm{MapScore}(tsk,\ acc,\ \mathrm{AccList,\ T_{curr},\ T_{cmpl},\ T_{deadline},}$ $\mathrm{Q_{task},\ Est_{Latency},\ Est_{Energy})}$
    
    \LeftComment{\% Compute base stats for scores}
    \Indent

    \State $ToGo(tsk) \leftarrow \sum_{i=1}^{N_{acc}}( \sum_{L \in Q_{task}(tsk)}\mathrm{Est_{latency}(i, L}))/N_{Acc} $     
    \State $Slack(tsk) \leftarrow \mathrm{T_{Deadline}}(tsk)\mathrm{-T_{curr}}$    
    \State $\mathrm{T_{queue}}(tsk) \leftarrow \mathrm{T_{curr} - T_{cmpl}}(tsk)$
    \State $ \mathrm{NextLayer}(tsk) \leftarrow Q_{task}(tsk).first $
    \State $ PrevAcc(tsk) \leftarrow Stack_{task}(tsk).acc $
    \EndIndent

    \LeftComment{\% Compute Urgency, Preference, and Starvation Scores}
    \Indent    
    \State $Score_{Urgency}(tsk) \leftarrow ToGo(tsk)/Slack(tsk)$
    \State $Score_{LatPref}(tsk,acc) \leftarrow \frac{\sum_{i=1}^{n} \mathrm{Est_{Latency}(NextLayer}(tsk), i)}{\mathrm{Est_{Latency}(NextLayer}(tsk), acc)}$
    \State $Score_{Starv}(tsk) \leftarrow \frac{\mathrm{T_{queue}}(tsk)}{\mathrm{ \sum_{i=1}^{N_{acc}} (Est_{Latency}(NextLayer}(tsk), i))/N_{acc}}$
    \EndIndent           
    
    \LeftComment{\% Compute the context-switch cost}
    \Indent
        \State $Cost_{switch}(tsk,acc) \leftarrow \frac{\mathrm{CswitchEnergy}(tsk, acc.\mathrm{prevTask}, acc)}{\mathrm{Est_{Energy}}(tsk,acc)}$ 
    \EndIndent
    \LeftComment{\% Compute Energy Score ($Score_{Energy}(tsk, acc)$)}
    \Indent
    \State $Pref_{Energy}(tsk,acc) \leftarrow \frac{\sum_{acc=1}^{N_{acc}} \mathrm{Est_{Energy}}(\mathrm{NextLayer}(tsk), acc)}{\mathrm{Est_{Energy}}(\mathrm{NextLayer}(tsk), acc)}$

    \State $Score_{Energy}(tsk,acc) \leftarrow $
    \Indent
    \State $Pref_{Energy}(tsk,acc) - Cost_{switch}(tsk,acc)$
    \EndIndent
    \EndIndent    
    
    \LeftComment{\% Compute total \scoreMetric \%}
    \Indent
    \State $\mathrm{MapScore} \leftarrow Score_{Urgency}(tsk)\cdot Score_{LatPref}(tsk,acc) $ 
    \Indent
    \Indent
    \State $ + \alpha\cdot Score_{Starv}(tsk) + \beta\cdot Score_{Energy}(tsk,acc)$
    \EndIndent
    \EndIndent    
    
    \Indent
    \textbf{return} MapScore
    \EndIndent 
    \EndIndent     
\end{algorithmic}

\end{algorithm}

\subsection{Urgency Score ($Score_{Urgency}$)}
\label{subsec:urgency_score}

Since we target RTMM ML applications, each inference request has a target deadline defined by the application and input streaming rates (i.e., FPS) from sensors or other data sources.
To facilitate scheduling while fully considering such real-time requirements, we propose the urgency score, $Score_{Urgency}$ (Line 7 in~\autoref{alg:mapscore}).
$Score_{Urgency}$ captures the capability of the underlying accelerators to satisfy the target deadline of an inference task ($tsk$).
It also considers the current system load and progress.

The urgency score is formulated for each inference task, considering the average latency across all accelerators.
To define $Score_{Urgency}$ of the $tsk$, we use $ToGo(tsk)$ and $Slack(tsk)$ described in ~\autoref{alg:mapscore} Lines 2-3.
$ToGo(tsk)$ quantifies the predicted remaining processing time of a model, and $Slack(tsk)$ quantifies the remaining time until the $tsk$ deadline.
Using $ToGo$ and $Slack$, $Score_{Urgency}$ quantifies the ratio of the predicted remaining processing time of a model ($ToGo(tsk)$) and the remaining time until the deadline ($Slack(tsk)$).
Based on the equation in Line 7, the urgency score increases either if we need a large amount of time to process an inference or if we have a short amount of available time for the inference request, which effectively models the urgency of the inference.

\subsection{Latency Preference Score ($Score_{LatPref}$)}
\label{subsec:latency_preference_score}

To identify the preference for accelerators for each layer (i.e., which accelerator provides lower latency and energy for each layer), \scheduler uses energy and latency estimations generated offline using a cost model or a simulator.
We first discuss the preference based on latency in this subsection and discuss how we can utilize energy preference later in~\autoref{subsec:enegy_score}.

To obtain the latency preference score, $Score_{LatPref}$ (~\autoref{alg:mapscore} Line 8), we first compute the fraction of latency on a target accelerator ($acc$) to the sum of latency on all accelerators that exist in the system.
This quantifies the significance of the latency on an $acc$ compared to all other accelerators, which is a lower-is-better metric.
By taking the inverse, as shown in Line 8 of~\autoref{alg:mapscore}, we obtain $Score_{LatPref}$ as a higher-is-better metric.

\subsection{Starvation Score ($Score_{Starv}$)}
\label{subsec:stravation_score}

Due to the workload and hardware heterogeneity, the latency of each layer has a high variance across hardware accelerators.
Such diverse latency can lead to the starving of light-weighted layers if the scheduler considers only the time margin to meet the target.
For example, if we expect operators A and B to take 1ms and 10ms for processing and both have 12ms until the deadline, a scheduler would schedule layer B first.
If such a situation is repeated, heavy-weighted operators can be continuously prioritized, leading to the starving of light-weighted operators.
Therefore, \scheduler uses a metric to measure the degree of starvation, Starvation score ($Score_{Starv}$). We provide the definition of $Score_{Starv}$ in Line 9 of~\autoref{alg:mapscore}.

$Starv$ quantifies the ratio of the wait time (i.e., queue time, $\mathrm{T_{queue}}$ in Line 4 of~\autoref{alg:mapscore}) and the latency of an operator estimated by a cost model such as MAESTRO~\cite{kwon2019understanding} and Timeloop~\cite{parashar2019timeloop}.
The score increases when a layer waits for a long time to be scheduled.
The wait time is divided by the estimated latency to provide a higher $Starv$ to light-weighted layers, which are less likely to be scheduled if we only consider the urgency and latency preference.

\subsection{Energy Score ($Score_{Energy}$)}
\label{subsec:enegy_score}

As many RTMM applications target battery-powered wearable (e.g., AR glasses) or mobile/edge (e.g., drone control) devices, energy is another important factor.
\scheduler optimizes energy consumption by identifying the most energy-efficient accelerator for each layer with the consideration of the context switch overhead.
$Score_{Energy}$ consists of two components: energy preference ($Pref_{Energy}$) and a context-switch cost function ($Cost_{switch}$), as Lines 10-13 in~\autoref{alg:mapscore} show.

Energy preference, $Pref_{Energy}$, is defined in a similar way to $Score_{LatPref}$; compute the significance of energy on an accelerator and take the inverse to obtain a higher-is-better metric. 
The context switch cost function, $Cost_{switch}$, computes the significance of the extra energy for context switching between two tasks on an accelerator.
The extra energy consists of the energy to fetch the activation of a new model from the DRAM and to flush that of the switched-out model to DRAM.
We define the energy score, $Score_{Energy}$ as $Pref_{Energy} - Cost_{switch}$ (~\autoref{alg:mapscore} Lines 12-13), which constructs a higher-is-better metric.

\subsection{Comprehensive Score: \scoreMetric}
\label{subsec:mapscore}


Combining all the scoring metrics we discussed, we define a comprehensive metric, \scoreMetric, as defined in Lines 14-15 of~\autoref{alg:mapscore}. 
\scoreMetric utilizes two parameters ($\alpha$ for starvation, $\beta$ for energy factor), which provide the adaptivity of \scoreMetric to individual systems and usage scenarios with different optimization goals.
We use \scoreMetric to drive scheduling decisions in \scheduler with $\alpha$ and $\beta$ optimization methodology we discuss next.

\subsection{\scoreMetric Parameter Optimization using \costFunction}
\label{subsec:parameter_optimization}

\begin{figure}[t]
    \centering
    \includegraphics[width=1\linewidth]{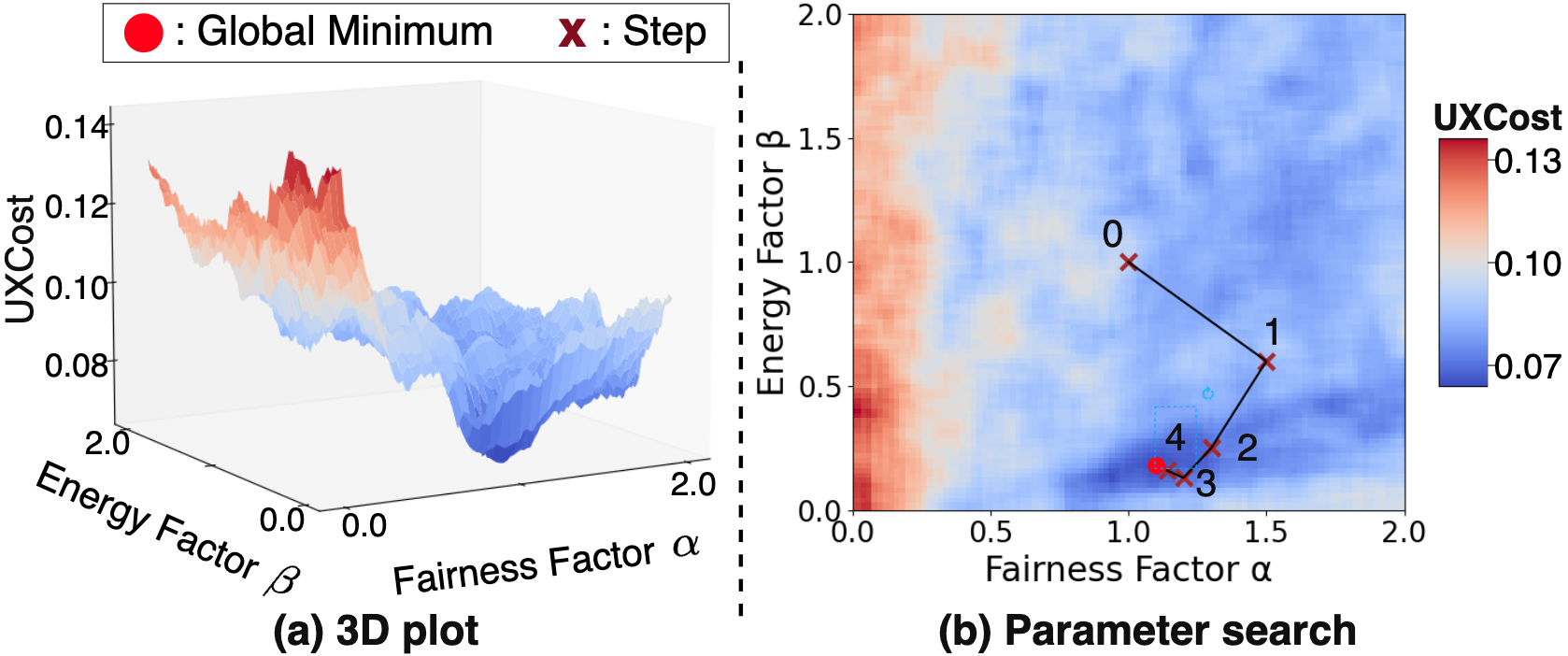}
    \caption{An example search space for scheduling score parameters and optimization steps. 
    } 
    \label{fig:parameter_optimization_motive}
\end{figure}



\begin{algorithm}[t]
\caption{\costFunction computation} \label{alg:uxcost}
\begin{algorithmic}[1]

\LeftComment{\textbf{Inputs}> \textbf{$\bm{Md}$}(A list of models in the workloads), $\bm{T_{exec}}$ (A time window where we run workloads)}
\LeftComment{\textbf{Output}> \costFunction($Md$, $T_{exec}$): \costFunction for $Md$ executed for $T_{exec}$
}
\State $\bm{Function} \ \mathrm{Compute\_}\costFunction(Md, T_{exec})$
\Indent
\LeftComment{\% Deadline (DL) violation rate and normalized energy for each model}
\State $Rate_{DLV}[len(Md)] \leftarrow \{0,\}$
\State $NormEnergy[len(Md)] \leftarrow \{0,\}$ 
\EndIndent
\LeftComment{\% $\forall m \in Md$, compute $Rate_{DLV}$ and $NormEnergy$ }
\Indent
\State $\text{\textbf{for}} \ m \ \text{\textbf{in}} \ Md \ \text{\textbf{do}}$
\Indent
\State $NormEnergy[m] \leftarrow \frac{\mathrm{Actual \  Total\ Energy\ Consumption}(m)}{\mathrm{Worst\ case\ Energy\ Consumption}(m)}$

\State $Rate_{DLV}[m] \leftarrow \frac{\mathrm{\#\ DL \ violated\ frames}(m, T_{exec})}{\mathrm{\#\ total\ frames}(m, T_{exec})}$

\EndIndent
\EndIndent
\LeftComment{\% Apply a small number to $Rate_{DLV}$ if no DL violation}
\Indent
\Indent
    \If {$\mathrm{\#\ DL \ violated\ frames}(m, T_{exec}) = 0$}
        \State $Rate_{DLV}[m] \leftarrow \frac{1}{2\cdot \mathrm{\#\ total\ frames}(m, T_{exec})}$
    \EndIf
\EndIndent
\State $\text{\textbf{end for}} $ 

\State $OverallRate_{DLV} \leftarrow \sum_{m \in Md} Rate_{DLV}[m]$ 
\State $OverallNormEnergy \leftarrow \sum_{m \in Md} NormEnergy[m]$  

\State $\mathrm{\costFunction} \leftarrow OverallRate_{DLV} \cdot OverallNormEnergy$
\State $\text{\textbf{return}} \ \costFunction$
\EndIndent
\end{algorithmic}

\end{algorithm}

\insertFigure{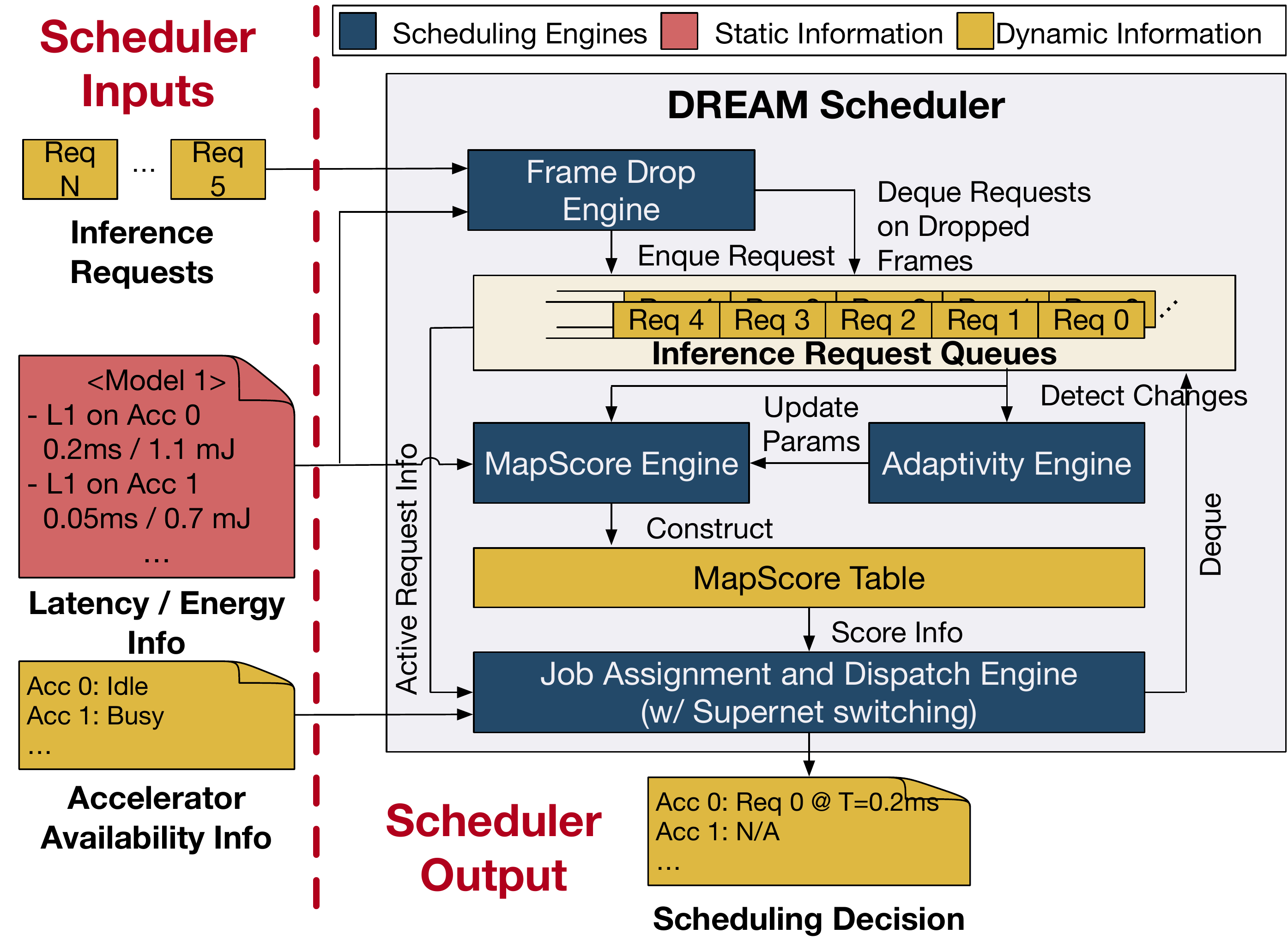}{An overview of the structure of \scheduler. Yellow boxes represent dynamic information, which is periodically generated (or when an event is detected) while running an RTMM workload. The red box has the latency and energy information for each layer in the models.}

\scoreMetric serves as a proxy metric during the scheduled time when actual metrics affecting overall user experience (e.g., deadline violation and energy) are unknown.
To utilize actual metrics viable to users (deadline violation and energy) for better scheduling decisions, we propose a methodology to provide feedback to \scoreMetric by adjusting starvation and energy factors ($\alpha$ and $\beta$).
To consider deadline violation and energy together, we construct an energy-delay product (EDP)-like metric, \costFunction, as defined in~\autoref{alg:uxcost}.
Like EDP, \costFunction is a lower-is-better metric that equally considers deadline violation rates and energy.
However, unlike EDP, \costFunction uses deadline violation instead of latency (delay) to reflect the deadline-driven nature of real-time applications.
Note that users can modify the formulation of \costFunction (e.g., squaring energy) to better align \costFunction with user-specific constraints.

~\autoref{alg:uxcost} describes a procedure to compute \costFunction, which investigates the deadline violation and energy consumption for each model in a workload.
Based on them, the procedure computes the deadline violation rate and normalized energy to the worst case (i.e. energy from worst layer-accelerator pairs) for each model within a given time window, $T_{exec}$ (e.g., 2 seconds).
Since the deadline violation rate of zero ($Rate_{DLV} = 0$) leads to \costFunction of zero, we apply small numbers based on the number of total frames shown in~\autoref{alg:uxcost} Lines 7-8.

\tobechecked{Minimizing \costFunction as a goal, we develop an iterative optimization method to optimize parameters ($\alpha$ and $\beta$) in \scoreMetric. It samples neighboring and distant parameter pairs, calculates the difference of the minimum two \costFunction pairs, and moves to the interpolated point. It repeats this process by decreasing the radius in the next step until the radius is below the threshold. 
This approach works well in \costFunction optimization space over $\alpha$ and $\beta$, which are well constrained and well defined, as ~\autoref{fig:parameter_optimization_motive} shows. 
With such well-conditioned, limited optimization space and quick convergence, we use $\alpha$ and $\beta$ for providing adaptivity to workload changes, which enhances the overall performance for dynamic workloads. We implement a lightweight online algorithm exploiting the quick convergence of the starvation and energy parameters, $\alpha$ and $\beta$. We discuss more detailed results in~\autoref{subsection:results}.}

\subsection{Why \scoreMetric is a Valid Metric for RTMM}

\begin{table}[t]
\caption{The consideration of RTMM challenges in previous schedulers and ours. \vspace{-1mm}}
\label{tab:rtmm-breakdown}
\resizebox{0.49\textwidth}{!}{%
\centering

\begin{tabular}{|c|c|c|c|c|c|c|} 
\hline

\multirow{3}{*}{\textbf{Scheduler}} 
    & \multicolumn{6}{c|}{\textbf{RTMM Challenges Consideration}} \\
    \cline{2-7} 
        & \multirow{2}{*}{Cascade} 
        & \multirow{2}{*}{Concurrent} 
        & \multirow{2}{*}{Real-time} 
        & \multicolumn{2}{c|}{\tobechecked{Workload dynamicity}}
        & \multicolumn{1}{l|}{\multirow{2}{*}{Energy}}  \\ 
        \cline{5-6} 
            & & & & \tobechecked{\ \ Task\ \ } & \tobechecked{Model} & \multicolumn{1}{l|}{} \\  
\hline
\multicolumn{1}{|l|}{Planaria, Veltair, MoCA} & \checkmark & \checkmark & \checkmark & & & \\ 
\hline
\multicolumn{1}{|l|}{\begin{tabular}[c]{@{}l@{}} Ours w/ MapScore \\ w/o param. optimization \end{tabular}} & \checkmark & \checkmark & \checkmark & & & \checkmark  \\ 
\hline
\multicolumn{1}{|l|}{\begin{tabular}[c]{@{}l@{}}Ours w/ MapScore \\ w/ param. optimization\end{tabular}}  & \checkmark & \checkmark & \checkmark & \checkmark & \checkmark & \checkmark \\
\hline
\end{tabular}
}
\vspace{-3mm}
\end{table}

~\autoref{tab:rtmm-breakdown} lists challenges of RTMM workloads mentioned in ~\autoref{sec:background} in columns and highlight which challenge \scoreMetric can handle, \tobechecked{compared to other dynamic schedulers for multi-task ML workloads, MoCA~\cite{moca}, Veltair~\cite{Veltair} and Planaria~\cite{Planaria}.}
Since all listed dynamic schedulers are deadline-aware dynamic schedulers for multi-accelerators, all schedulers can deal with ML model cascade, ML model concurrency, and real-time timing requirements.
However, prior schedulers employ fixed scheduling algorithms for all the usage scenarios, which lead to insufficient adaptivity to the RTMM workload dynamicity.
In contrast, \scheduler can adapt to various usage scenarios in RTMM applications utilizing the \scoreMetric parameter ($\alpha$ and $\beta$) online optimization algorithm we discussed in~\autoref{subsec:parameter_optimization}.
%


\section{\scheduler Scheduler}
\label{sec:scheduler}



Using the scoring metrics we define in~\autoref{sec:score}, we implement \scheduler scheduler, which handles all the new challenges from RTMM workloads discussed in~\autoref{sec:RTMM_workloads}. 
We illustrate the overall structure of \scheduler in~\autoref{fig:SchedulerOverview} with the internal flow within the scheduler.
As inputs, \scheduler receives (1) inference requests, which would be periodically generated based on streamed input data, (2) latency and energy information for each layer for each accelerator in the system generated offline using a cost model or a simulator, and (3) accelerator availability information from hardware to determine accelerators that can accept new inference jobs.
As output, \scheduler generates the scheduling decision, which includes information on the scheduled inference jobs and their target accelerators with the dispatch time. 

\subsection{Scheduling Flow}
\label{label:scheduler_engines}

\scheduler consists of four main software components, labeled as scheduling engines in~\autoref{fig:SchedulerOverview}: frame drop engine, MapScore calculator, adaptivity engine, and job assignment / dispatch engine. 
When an input inference request arrives, the frame drop engine checks the status of the inference request queue to determine acceptance of the inference request.
After injecting the request to an inference request queue, the Adaptivity Engine checks changes in workloads and triggers the starving and energy parameter tuning described in~\autoref{subsec:parameter_optimization} if changes are detected.
The MapScore Engine computes the \scoreMetric for requested inference on all the accelerators in the target ML system using the latency and energy information generated from an offline cost model or simulator.
The calculated \scoreMetric is stored in the MapScore table.
Finally, the job assignment/dispatch engine drives the scheduling decision based on \scoreMetric in the MapScore table, current accelerator availability information (i.e., which accelerator is busy and idle), and current inference requests in the request queue.

The flow includes four core optimizations we implement in \scheduler: Light-weighted score metric computation, Light-weighted Adaptivity Engine (with online parameter tuning), Smart Frame Drop, and Supernet Switching. 

\subsection{Frame Drop Engine}
\label{subsec:smart_frame_drop}

Although many previous works explored frame drop techniques~\cite{skip-over, combining_nm_hard_scheduling, m_k_firm_overload_management} to adjust overall system load to a healthy region, they are not tailored for RTMM.
As a result, such works fail to capture key challenges of RTMM workloads: (1) model dependency from cascaded models and (2) task, model, and operator level dynamicity.
We clarify key requirements for a frame drop methodology for RTMM workloads.

\betterparagraph{Requirement 1: Considering Model Dependency} 
 When executing cascaded models as an ML pipeline, dropping a frame in a model naturally leads to another frame drop of the next models in the dependency chain. As more than one model can rely on the results of the precedent model, the impact of frame drop on an early model shouldn't be underestimated.

\betterparagraph{Requirement 2: Exploiting Constrained Dynamicity} 
Past works rely on some statically defined parameters and mechanisms~\cite{skip-over, combining_nm_hard_scheduling, m_k_firm_overload_management} to handle dynamic workloads, because we cannot predict workloads in the general-purpose computing context. 
However, on the RTMM workloads, the workloads are not completely random, as all the possible tasks, models, and operators can be identified by the workload specifications (e.g., In a hand-tracking ML pipeline of an AR workload, we know there are only two choices after a hand detection model: not launching or launching a hand tracking model).
This opens up RTMM-specific optimization opportunities, which we exploit in the frame drop engine.

\subsubsection{Smart Frame Drop Mechanism.}
\label{subsubsec:smart_frame_drop}

\begin{figure}[t]
    \centering
    \includegraphics[width=0.9\linewidth]{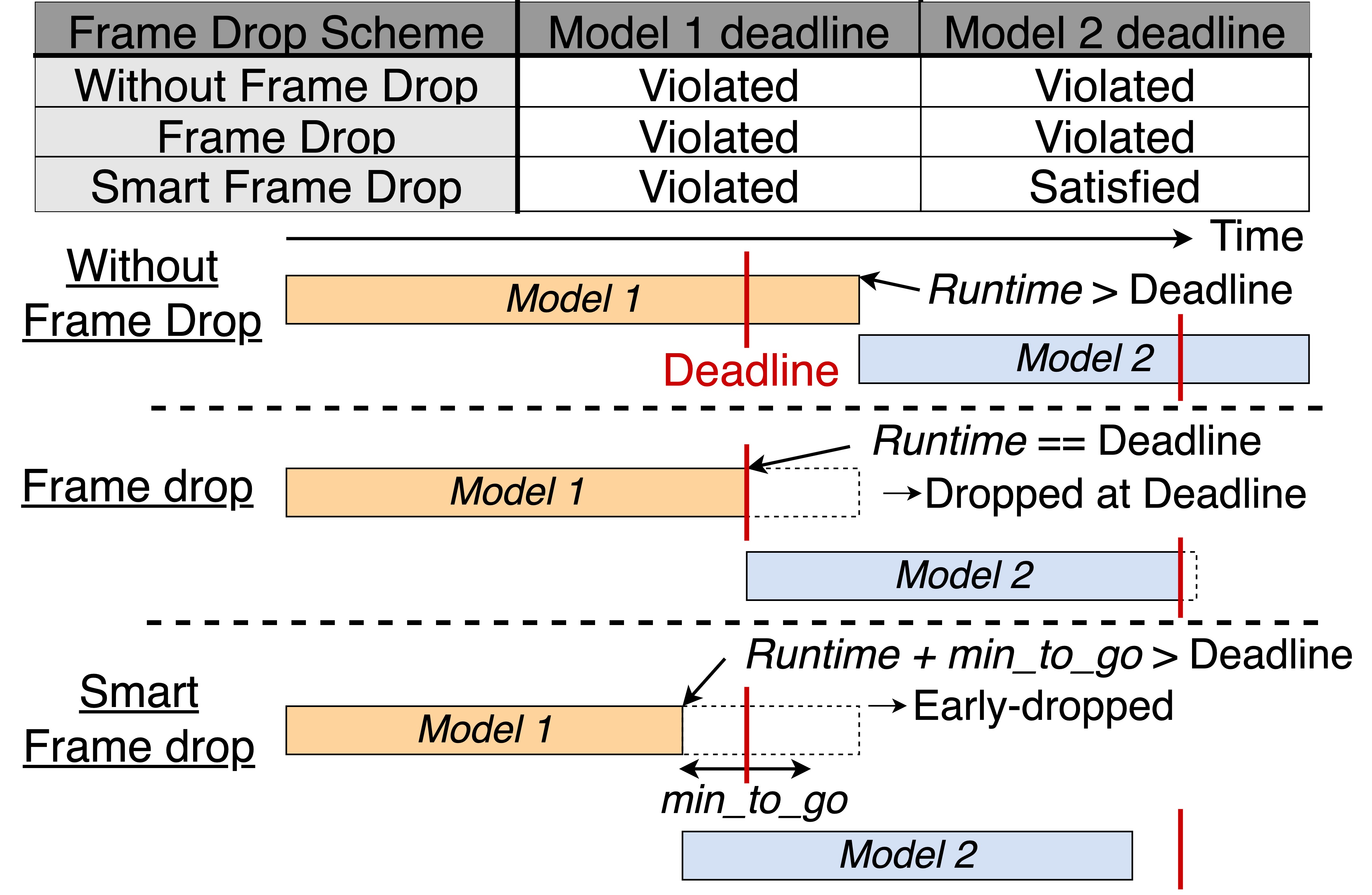}
    \caption{Comparison of smart frame drop with other frame drop schemes.\vspace{-2mm}}
    \label{fig:framedrop}
\end{figure}

Based on the requirements, we develop an RTMM-oriented frame drop mechanism,  smart frame drop.
In addition to our insights on the RTMM workload dynamicity, 
Smart frame drop exploits the predictability of latency of each layer in ML accelerators~\cite{parashar2019timeloop,kwon2019understanding} by utilizing the layer-wise latency and energy information input to \scheduler as shown in~\autoref{fig:SchedulerOverview}.
%
The latency and energy information enables the frame drop engine to estimate the remaining time until the deadlines of a model, which is a part of key information for determining smart frame drop. 
~\autoref{fig:framedrop} illustrates our frame drop mechanism, smart frame drop, and compares it against other frame drop mechanisms.
Unlike other methods, smart frame exploits RTMM-specific characteristics by identifying the next workloads (worst and best cases based on the constrained dynamicity) and utilizing pre-computed latency information (the predictability of latency for each layer).
Using such information, the smart frame drop predicts the possibility of deadline violations before reaching deadlines and proactively drops frames if the deadline is unlikely to be met.
As a result, smart frame drop reduces overall deadline violations by providing more time to other models as shown in~\autoref{fig:framedrop}, which is not possible in traditional frame drop methodologies without RTMM-specific workload knowledge.

The frame drop engine is triggered each time a new scheduling decision needs to be made in the job assignment and dispatch engine.
The smart frame drop only drops a frame only if all the following four conditions are met.

\betterparagraph{Condition 1: Deadline Violation Likelihood}
Jobs expected to miss deadlines should be the frame drop target.
To identify such targets, the smart frame drop mechanism checks the following condition: $minimum\_to\_go > Slack$ where $minimum\_to\_go$ refers to the minimum remaining time until completion, assuming that the best-latency accelerator is used for each layer without any context switching, and $Slack$ refers to the remaining time until the deadline. 

\betterparagraph{Condition 2: Multi-model Violation}
More than one active job in accelerators should be expected to violate deadlines.
This prevents redundant frame drops when completing the current inference late does not affect other models' deadlines.

\betterparagraph{Condition 3: Dependency-free}
Only the last model in a dependency chain (i.e., an ML pipeline), which does not have any other models that depend on the model, can be a frame drop target.
This corresponds to Requirement 1 in~\autoref{subsec:smart_frame_drop}.

\betterparagraph{Condition 4: Maximum Frame Drop Rate}
To prevent excessive frame drops on a specific model, the engine bounds the maximum frame drop rate over a specified frame window length.
The rate is a configurable parameter, and by default, smart frame drop allows up to 2 drops per 10 frames.


The frame drop engine identifies the frame with the highest $minimum\_to\_go/slack$ among all the frames meeting all four conditions and drops the frame if exists.
We interpret the frame drop as deadline violations (completion time = $\infty$) and consider its impact in \costFunction defined in~\autoref{alg:uxcost}.

\subsection{\scoreMetric Engine}
\label{subsec:opt_scoring}

To tackle the dynamicity of RTMM discussed in~\autoref{sec:dynamicity}, we implement a light-weighted online dynamic scheduling mechanism, \scoreMetric Engine, which computes \scoreMetric of the top layers in each inference request queue based on~\autoref{alg:mapscore}.
\scoreMetric Engine uses pre-computed latency and energy for each accelerator using an offline cost model or simulator.
This exploits determinism of latency and energy in accelerators once an inference launches if all data are loaded~\cite{parashar2019timeloop, kwon2019understanding}, which is a main difference between dense ML accelerators and other hardware options such as GPUs.

\subsection{Adaptivity Engine}
\label{subsec:adpativity_engine}

When the target workload changes, the optimal starvation and energy factors ($\alpha$ and $\beta$) in \scoreMetric also change. 
Adaptivity Engine, illustrated in ~\autoref{fig:SchedulerOverview}, detects the workload changes by tracking the inference model list and triggers the starvation and energy factor optimization discussed in~\autoref{subsec:parameter_optimization}. 
In addition to the workload change scenario, Adaptivity Engine provides adaptivity to system load changes by tracking overall deadline violation rates and frame drop rates, without adding extra latency to the end-to-end latency. It continuously tests a small number of pairs ($\alpha$, $\beta$) around the current value for a short time window and makes a move to a pair that provides the lowest \costFunction value. That is, \scheduler keeps generating valid schedules while gradually optimizing its internal strategy to a new environment.

\subsection{Job Assignment and Dispatch Engine} 
\label{subsec:job_assignment_and_dispatch_engine}

\begin{figure}[t]
    \centering
    \includegraphics[width=0.95\linewidth]{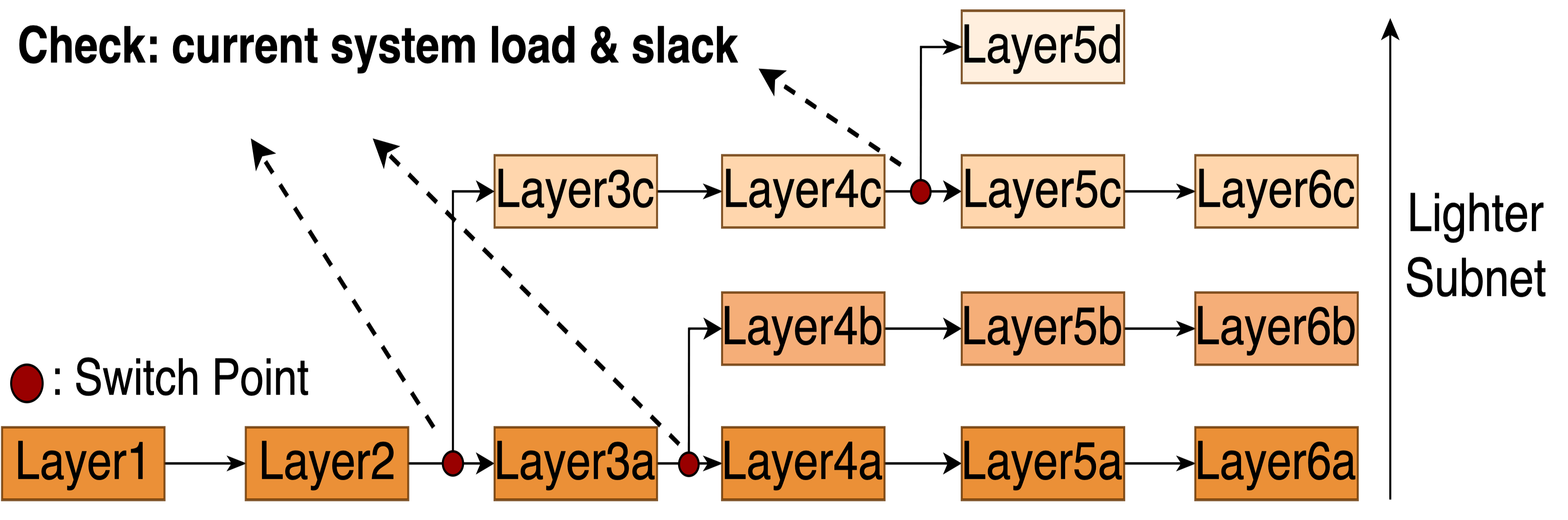}
    \caption{An example of Supernet switching that deploys lighter subnets based on the system load and slack (remaining time until deadline - expected completion time) during execution.\vspace{-2mm}}
    \label{fig:supernet}
\end{figure}

By default, Job Assignment and Dispatch Engine draw a scheduling decision by selecting the layer-accelerator pair with the highest \scoreMetric.
Optionally, Supernet switching can be activated, which uses a technique explored by previous work~\cite{once-for-all} in the ML algorithm domain.

\subsubsection{Supernet-switching.}
\label{subsubsec:supernet_switching}

The Supernet weight-sharing technique has been used for training a model once and optimizing the model for different deployment scenarios~\cite{once-for-all}.
Unlike the previous work targeting the static selection of sub-net instances within a Supernet, we explore an RTMM-tailored use case, which is the online sub-net instance switching, as a future-proof form of Supernet switching.
This approach helps the system to adaptively decrease the overall system load by deploying lighter model instances, which contributes to better user experiences as it reduces overall deadline violation rates.

~\autoref{fig:supernet} shows an example of a Supernet-based model with runtime model instance switching.
When the job assignment and dispatch engine generates a scheduling decision on a Supernet layer followed by a switch point, it decides which branch to take by estimating the possibility of whether the current workload can meet the deadline or not.
If it cannot meet the deadline, the job assignment and dispatch engine switch to a lighter variant of the Supernet model.

\subsubsection{\tobechecked{Impact on Accuracy.}}
To prevent accuracy degradation, we carefully select Supernet variants that provide equivalent or greater accuracy than similar-sized or larger networks.
For example, Once-for-all~\cite{once-for-all} variant ofa-s7edge-41 has 96MFLOPs with 73.1\% of Top1-accuracy while a fined-tuned MobileNetV3 (at s7edge, similar FLOPs) and MobileNetV2 provides 70.4\% and 71.8\%-Top1-accuracy, respectively.
All the Supernet variants we use (ofa-s7edge) provide higher accuracy with fewer FLOPs than similar-sized models.
Note that the online sub-net instance switching implemented in \scheduler does not further degrade the model accuracy beyond the baseline static methods~\cite{once-for-all}.
\section{Evaluation}
\label{sec:evaluation}

We evaluate \scheduler against three baseline dynamic schedulers, FCFS (first-come-first-served), Veltair~\cite{Veltair}, and Planaria~\cite{Planaria} using five realistic RTMM workload scenarios based on the industry works~\cite{xrbench, trailnet}. 

\begin{table}[t]
\caption{Evaluated accelerator hardware settings \vspace{-1mm}}
\label{tab:target_hw_settings}
\scriptsize
\centering
\resizebox{0.45\textwidth}{!}{%
\begin{tabular}{|c|l|l|}
\hline
\multicolumn{1}{|c|}{
\textbf{\begin{tabular}[c]{@{}c@{}}Size \\ (\# of PE)\end{tabular}}} 
& \multicolumn{1}{c|}{\textbf{Style}} 
& \multicolumn{1}{c|}{\textbf{\begin{tabular}[c]{@{}c@{}}Dataflow \\ (PE partitioning)\end{tabular}}} \\ \hline
\multirow{4}{*}{4K} 
  & \multirow{2}{*}{Homogeneous}  
    & 2 WS (2K each) \\ 
    \cline{3-3}
    & & 2 OS (2K each)  \\
    \cline{2-3}
  & \multirow{2}{*}{Heterogeneous} 
    & 1 WS (2K) + 2 OS (1K each) \\
    \cline{3-3}
    & & 1 OS (2K) + 2 WS (1K each) \\
    \hline
\multirow{4}{*}{8K} 
    & \multirow{2}{*}{Homogeneous} 
      & 2 WS (4K each) \\ 
      \cline{3-3}
      & & 2 OS (4K each) \\ 
      \cline{2-3}
    & \multirow{2}{*}{Heterogeneous} 
      & 1 WS (4K) + 2 OS (2K each) \\ 
      \cline{3-3}
      & & 1 OS (4K) + 2 WS (2K each) \\
      \hline
\end{tabular}%
}
\end{table}
\begin{table}[t]
\caption{Evaluated Real-time workload scenarios with their FPS targets (FPS) and dependencies (Dep). HD, KS, and FD refer to hand detection, keyword spotting, and face detection models.}
\label{tab:workload_scenarios}
\resizebox{0.48\textwidth}{!}{%
\centering
\begin{tabular}{|c|l|l|l|l|} 
\hline
\multicolumn{1}{|c|}{\textbf{Scenario}}                      & \multicolumn{1}{c|}{\textbf{Application}} 
& \multicolumn{1}{c|}{\textbf{Model}} 
& \multicolumn{1}{c|}{\textbf{FPS}} 
& \multicolumn{1}{c|}{\textbf{Dep.}}  \\ 
\hline

\multirow{6}{*}{\shortstack[lb]{VR \\ Gaming}}
  & Gaze Estimation & FBNet-C~\cite{fbnet} & 60 & \\
\cline{2-5}
  & Hand Detection & SSD\_MobileNetV2~\cite{SSD} & 30 & \\
  \cline{2-5}
  & Pose Estimation & HandPoseNet~\cite{handposenet} & 30 & HD \\
  \cline{2-5}
  & Context understanding & Once-for-all~\cite{once-for-all} & 30 & \\
  \cline{2-5}
  & Keyword Spotting & KWS\_res8~\cite{kws-res26} & 15 & \\
  \cline{2-5}
  & Translation & GNMT~\cite{gnmt} & 15 & KS \\
\hline
\multirow{3}{*}{\shortstack[cb]{AR \\ Call}}
  & Keyword Spotting & KWS\_res8~\cite{kws-res26} & 15 & \\
  \cline{2-5}
  & Translation & GNMT~\cite{gnmt} & 15 & KS \\
  \cline{2-5}
  & Context understanding & SkipNet~\cite{SkipNet} & 30 & \\
\hline
\multirow{3}{*}{\shortstack[lb]{Drone \\ (Outdoor)}} 
  & Object Detection & SSD\_MobileNetV2~\cite{SSD} & 30 & \\
  \cline{2-5}
  & Outdoor Navigation & TrailNet~\cite{trailnet} & 60 & \\
  \cline{2-5}
  & Visual Odometry & SOSNet~\cite{sosnet} & 60 & \\
\hline
\multirow{4}{*}{\shortstack[lb]{Drone \\ (Indoor)}}
  & Object Detection & SSD\_MobileNetV2~\cite{SSD} & 30 & \\
  \cline{2-5}
  & Indoor Navigation & RAPID\_RL~\cite{RAPID-RL} & 60 & \\
  \cline{2-5}
  & Obstacle Detection  & SOSNet~\cite{sosnet} & 60 & \\
  \cline{2-5}
  & Car Classification  & GoogLeNet-car~\cite{googlenet-car} & 60 & \\
\hline
\multirow{5}{*}{\shortstack[lb]{AR\\Social\\Interaction}} 
  & Depth Estimation & FocalLengthDepth~\cite{focallengthdepth} & 30 & \\
  \cline{2-5}
  & Action Segmentation & ED-TCN~\cite{tcn} & 30 & \\
  \cline{2-5}
  & Face Detection  & SSD\_MobileNetV2~\cite{SSD}  & 30 & \\
  \cline{2-5}
  & Face Verification & VGG-VoxCeleb~\cite{voxceleb} & 30 & FD \\
  \cline{2-5}
  & Context Understanding & Once-for-all~\cite{once-for-all} & 30 & \\
\hline
\end{tabular}
}
\end{table}

\begin{table}[t]
\caption{\scheduler configuration used in evaluation}
\label{tab:scheduler_configuration}
\resizebox{0.48\textwidth}{!}{%
\centering
\begin{tabular}{|l|c|c|c|} 
\hline
\multicolumn{1}{|c|}{\begin{tabular}[c]{@{}c@{}}\textbf{\scheduler}\\\textbf{Configurations}\end{tabular}} & \begin{tabular}[c]{@{}c@{}}\textbf{Dynamic Score}\\\textbf{Parameter Optimization}\end{tabular} & \begin{tabular}[c]{@{}c@{}}\textbf{Smart}\\\textbf{Frame Drop}\end{tabular} & \begin{tabular}[c]{@{}c@{}}\textbf{Supernet}\\\textbf{Switching}\end{tabular}  \\ 
\hline
\dyscore                           & \cmark                                                                             &                &                   \\
\hline
\smartdrop                         & \cmark                                                                             & \cmark                &                    \\
\hline
\full                               & \cmark                                                                             & \cmark                & \cmark       \\
\hline
\end{tabular}
}
\end{table}


\subsection{Evaluation Setting}
\label{subsec:eval_setting}

\betterparagraph{Target hardware}
~\autoref{tab:target_hw_settings} lists the hardware systems we evaluated.
To show the efficacy of \scheduler on various hardware platforms with accelerators, we vary the size of the total number of processing elements (PEs), 4K and 8K, and the dataflows, homogeneous and heterogeneous dataflow with weight-stationary (WS) and output-stationary (OS) dataflows.
The WS and OS dataflows are inspired by NVDLA~\cite{nvdla} and Shidiannao~\cite{du2015shidiannao} with further tile size optimizations. 
We also vary the number of PEs across sub-accelerator instances as shown in~\autoref{tab:target_hw_settings} to model various systems.
For all accelerators, we assume 8 MiB of on-chip shared SRAM with 90 GB/s of off-chip bandwidth, running at a 700MHz clock frequency.


\betterparagraph{Evaluated workload scenarios}
We construct five RTMM scenarios, which are shown in ~\autoref{tab:workload_scenarios}. 
\ar, \vr and \texttt{AR\_Call} are based on XRBench~\cite{xrbench}, and \texttt{Drone\_Outdoor} and \texttt{Drone} \texttt{\_Indoor} are based on TrailMAV~\cite{trailnet}.
\ar and \vr have ML pipelines with control and data dependency as "Dep." in~\autoref{tab:workload_scenarios} shows. By default, we activate the dependent workload with 50\% of probability. For Supernet switching, we used four Once-for-All~\cite{once-for-all} model variants for the (visual) context understanding workload. We use four different sub-networks of the Supernet model with weight sharing. 
We also include operator-level dynamic networks, such as SkipNet~\cite{SkipNet} and RAPID\_RL~\cite{RAPID-RL} with early-exit branches. We apply the branch exit probability that each work presented (e.g., the probability of 50\% for each block for SkipNet, which reported over 72\% of Top-1 accuracy on ImageNet). 
In addition to dynamicity, the workloads model concurrent ML pipelines. 
For the outdoor drone scenario, we adopt the workload presented in TrailMAV~\cite{trailnet}. For the indoor drone scenario, we replace the navigation model in TrailMAV~\cite{trailnet} with RAPID\_RL~\cite{RAPID-RL}, which is tailored for indoor environments. We also use GoogLeNet-car to target in-door parking enforcement use scenarios. For \ar, we create speaker identifying ML pipeline based on \cite{voxceleb} and use its VGG-based model for active speaker verification.



\begin{figure*}[t]
    \centering
    \includegraphics[width=1.0\linewidth]{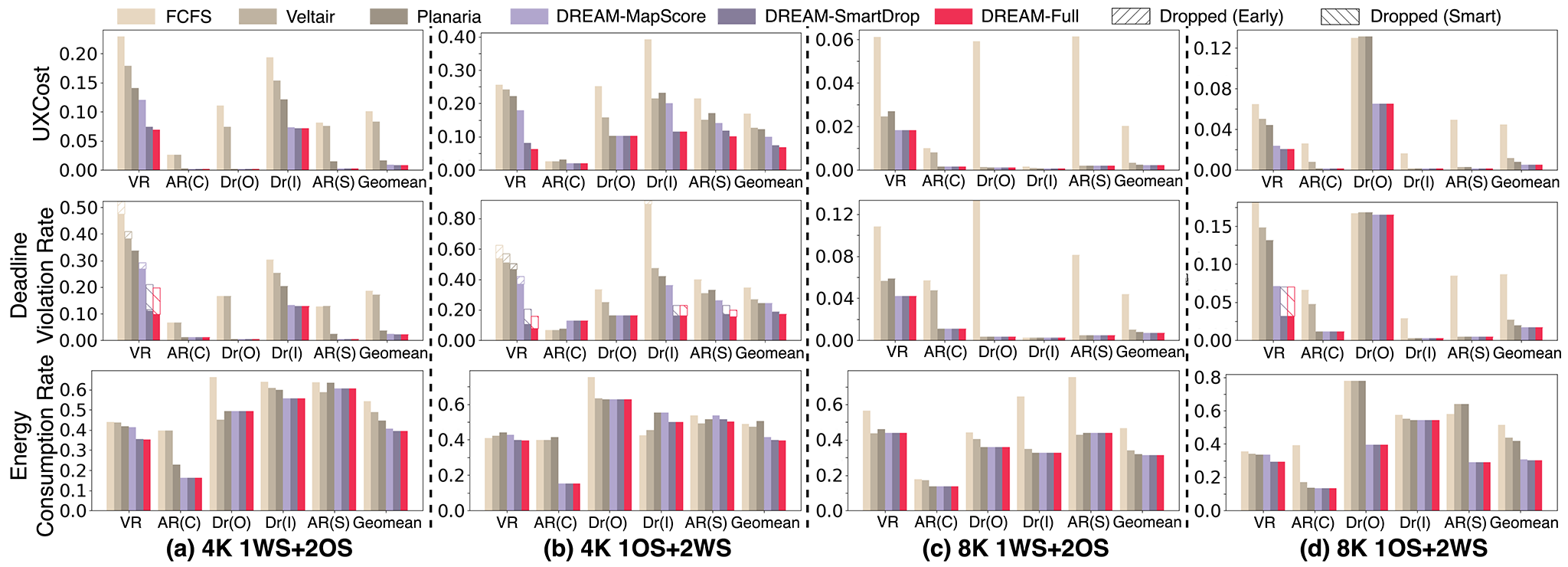}
    \caption{\costFunction, Deadline Violation Rate, Energy Consumption evaluation on different workloads and target hardware. Energy consumption is normalized to the maximum possible energy consumption of the target hardware. x-axis denotes workload scenarios.\vspace{0mm}}
    \label{fig:uxcost}
\end{figure*}

\begin{figure}[t]
    \centering
    \includegraphics[width=1.0\linewidth]{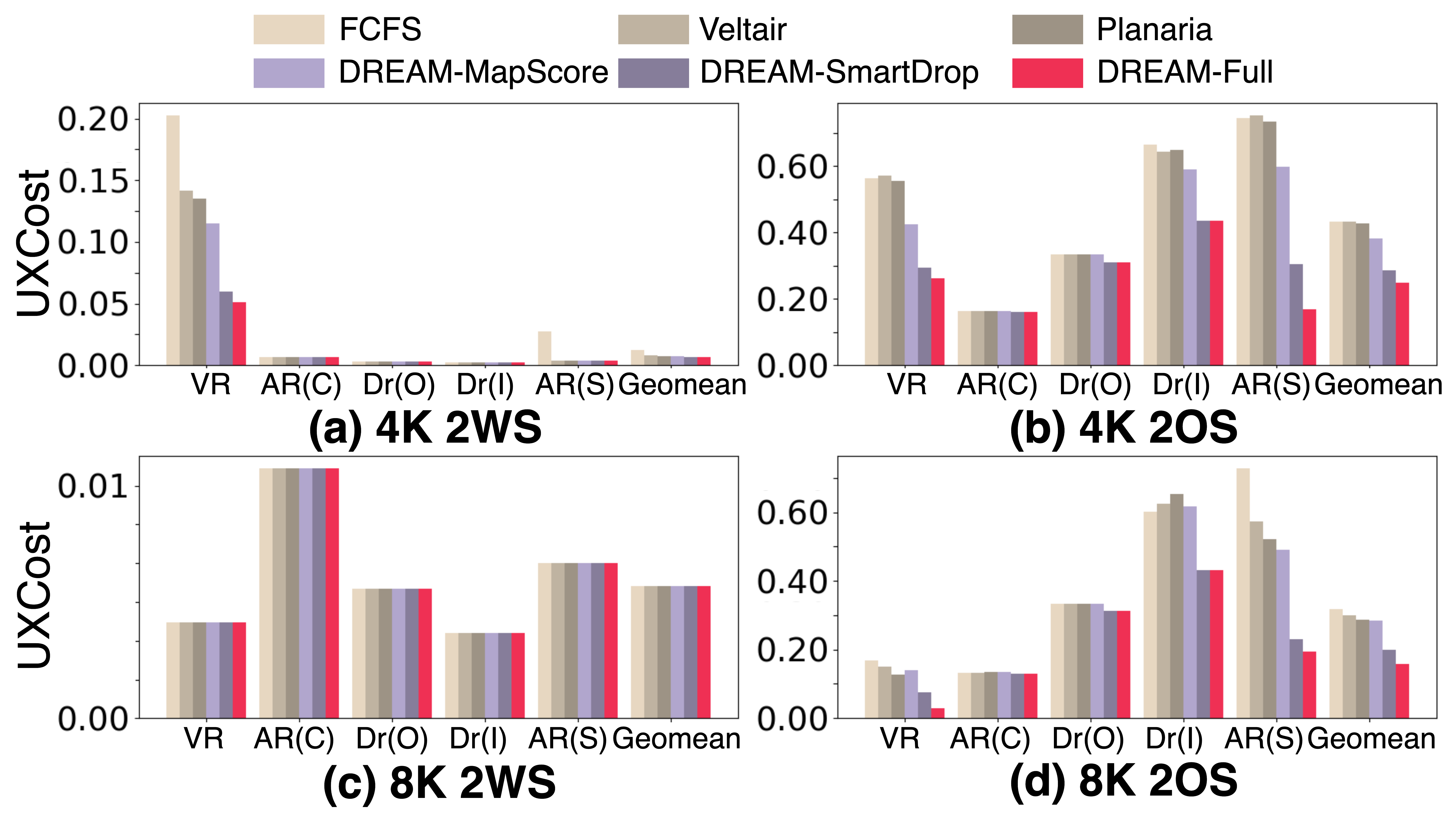}
    \caption{\costFunction of homogeneous target hardware settings.\vspace{-2mm}} 
    \label{fig:uxcost_homo}
\end{figure}

\betterparagraph{Schedulers}
We compare our \scheduler with three dynamic scheduling algorithm baselines with different scheduling granularities and strategies:
\begin{enumerate}
    {\item \textbf{First-Come-First-Served (FCFS)}~\cite{clockwork, Nexus}: serves the oldest request in the queue immediately if there is a resource available in the granularity of the model.}
    {\item \textbf{Veltair}~\cite{Veltair}: schedules threshold-based layer-block which groups consecutive layers to prevent scheduling conflicts.}
    {\item \textbf{Planaria}~\cite{Planaria}: spatially co-locates multiple DNNs by dynamically partitioning the compute resources layerwise based on timing requirements and resource demands.}
\end{enumerate}
\tobechecked{Note that Veltair is a framework performing both scheduling and compilation that targets a homogeneous CPU cluster. We model their layer-blocking scheme and scheduler. For Planaira, note
that it is based on the hardware-software co-design, whereas we model the scheduling component.
For simplicity, we refer to the schedulers of Veltair and Planaria as Veltair and Planaria in this section.} To analyze the impact of the smart frame drop and Supernet switching, we evaluate the two variants of \scheduler along full \scheduler, as listed in~\autoref{tab:scheduler_configuration}.
\dyscore performs \scoreMetric parameter optimization along the score metric-driven job assignments, but without smart frame drop and Supernet switching.
\smartdrop enables the smart frame drop upon \dyscore. We set the maximum frame drop rate at 20\%. \full includes all optimization of \scheduler, which indicates \smartdrop with Supernet switching.

\betterparagraph{Evaluation metric} We use \costFunction as our main evaluation metric, which is a product of the deadline violation rate and the energy consumption rate. As discussed in~\autoref{subsec:parameter_optimization}, \costFunction is a comprehensive metric that considers two key aspects that affect the overall user experience: the deadline violation rate and the energy consumption, which are aligned with the energy-delay product (EDP) used in non-real-time systems.

\betterparagraph{Latency and Energy Estimation} 
We use MAESTRO~\cite{kwon2019understanding} cost model to obtain latency and energy, which reported near 96\% accuracy for estimations.

\subsection{Results and Discussions}
\label{subsection:results}

\betterparagraph{\scheduler significantly outperforms baselines} We compare variants of \scheduler listed in~\autoref{tab:scheduler_configuration} against baseline schedulers in~\autoref{fig:uxcost} and~\autoref{fig:uxcost_homo}. On average across all workload scenarios and hardware settings, \scheduler decreases overall \costFunction by 32.1\% and 50.0\% against Planaria~\cite{Planaria} and Veltair~\cite{Veltair}. 
In particular, we observe high improvements for the \ar scenario with 4K PEs(1WS+2OS), reducing \costFunction by 80.8\% compared to Planaria, and \texttt{Drone\_Outdoor}  scenario with 4K PEs(1WS+2OS), reducing \costFunction by 97.6\% compared to Veltair, as shown in~\autoref{fig:uxcost} (a).
For the deadline violation rate decrease, considering heterogeneous hardware by the preference score and heterogeneous workload by the starvation score efficiently handled relatively heavy workloads.
For reducing energy consumption, the energy score helped the energy-aware scheduling optimization, while other baselines do not consider energy.

\betterparagraph{\scheduler's HW-heterogeneity-aware scheduling is effective for heterogeneous HW} 
The overall \costFunction gap against \scheduler in the heterogeneous environment~\autoref{fig:uxcost} is larger than in homogeneous results in~\autoref{fig:uxcost_homo}, by $2.20\times$ for Veltair and $1.26\times$ for Planaria.
The results imply that a holistic consideration of hardware heterogeneity is important for schedulers; only \scheduler considers accelerator size, shape, and dataflow entirely, which distinguishes it from the baselines.
%


\betterparagraph{Scheduling is important when the computing power is constrained}
When a system has abundant compute resources, the impact of scheduling can diminish as the results in~\autoref{fig:uxcost_homo} (c). However, as many RTMM workloads are expected to be supported on wearable (e.g., AR glasses) or mobile/edge (e.g., drone) devices with more ML models, constrained compute resources environments are expected to be common cases. For such environments, \scheduler demonstrates its strengths. For example, in the 4K 1WS+2OS setting, \scheduler on average provides 
decrease in \costFunction by 47.5\% and 89.9\% compared to Planaria and Veltair, respectively.
Unlike baselines, \scheduler implements a dynamic scheduling algorithm that utilizes runtime information to adapt to dynamic system load changes, which enables superior results compared to baseline in resource-constrained settings.


\betterparagraph{\scheduler provides considerable energy reduction}
Both Veltair and Planaria do not optimize for energy consumption. In contrast, \scheduler improves energy consumption even under severe resource constraints based on its adaptive workload adjustment schemes. 
For example, \scheduler decreased energy consumption by 62.9\% and 61.4\% compared to Planaria and Veltair, respectively, for \texttt{AR\_Call} on 4K 1OS+2WS system.
Although \scheduler resulted in a $1.88\times$ and $1.66\times$ higher deadline violation rate, overall \costFunction decreased by 27.4\% and 38.3\% respectively, because the baselines are mainly optimized for deadlines.
\scheduler has the flexibility to explore the balance between deadline and energy by updating the fairness and energy factors ($\alpha$ and $\beta$) depending on the application and system requirements.
Another example is \ar executed on 8K 1OS+2WS where \scheduler reduces the energy by 54.6\% against the baselines, on average.
Such data show that our energy score is an effective way to guide energy optimization in a scheduler.


\begin{figure}[t]
    \centering
    \vspace{-3.5mm}
    \includegraphics[width=1.0\linewidth]{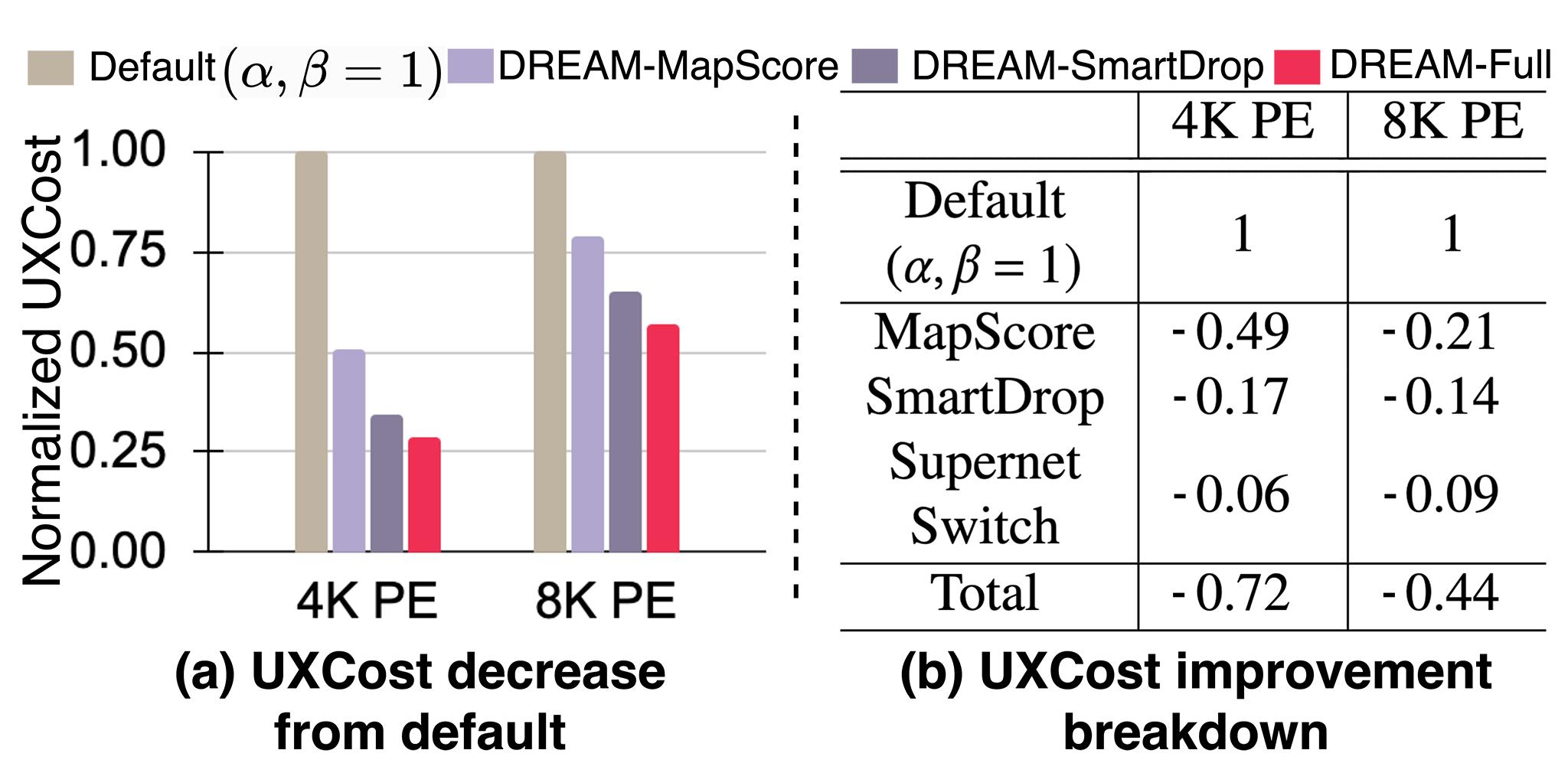}
    \caption{\vr and \ar geomean \costFunction improvement breakdown for each optimization method. \vspace{-4mm}}
\label{tab:uxcost-breakdown}
\end{figure}

\betterparagraph{Each optimization component of \scheduler is effective}
~\autoref{tab:uxcost-breakdown} shows the geometric mean performance improvement breakdown against baseline \scoreMetric with fixed $\alpha, \beta = 1$, as enabling \scoreMetric optimization for \scoreMetric (\dyscore), enabling smart frame drop (\smartdrop), and enabling Supernet switch (\full) for \vr and \ar, which contain Supernet models.
The result shows the efficacy of all three optimizations.
The parameter optimization of \scoreMetric alone shows  \costFunction decrease by 49.2\% for 4K PE and 21.0\% for 8K PE case from when \scoreMetric's parameter $\alpha, \beta$ are fixed to 1.
Enabling Smart frame drop on top of \scoreMetric optimization shows about 16.5\% (4K) and 13.8\% (8K). 
Supernet Switch further decreases \costFunction by 6-9\% across the configurations.

Note that smart frame drop and Supernet switching do not add extra latency because their latency is small and overlaps with the actual execution of workloads. In addition, they do not pose a negative impact on compute resource-sufficient scenarios. For example, ~\autoref{fig:uxcost_homo} (c) shows no difference among \dyscore, \smartdrop, and \full, which indicates negligible overhead of those two techniques in compute resource-sufficient scenarios.

\begin{figure}[t]
    \centering
    \includegraphics[width=1\linewidth]{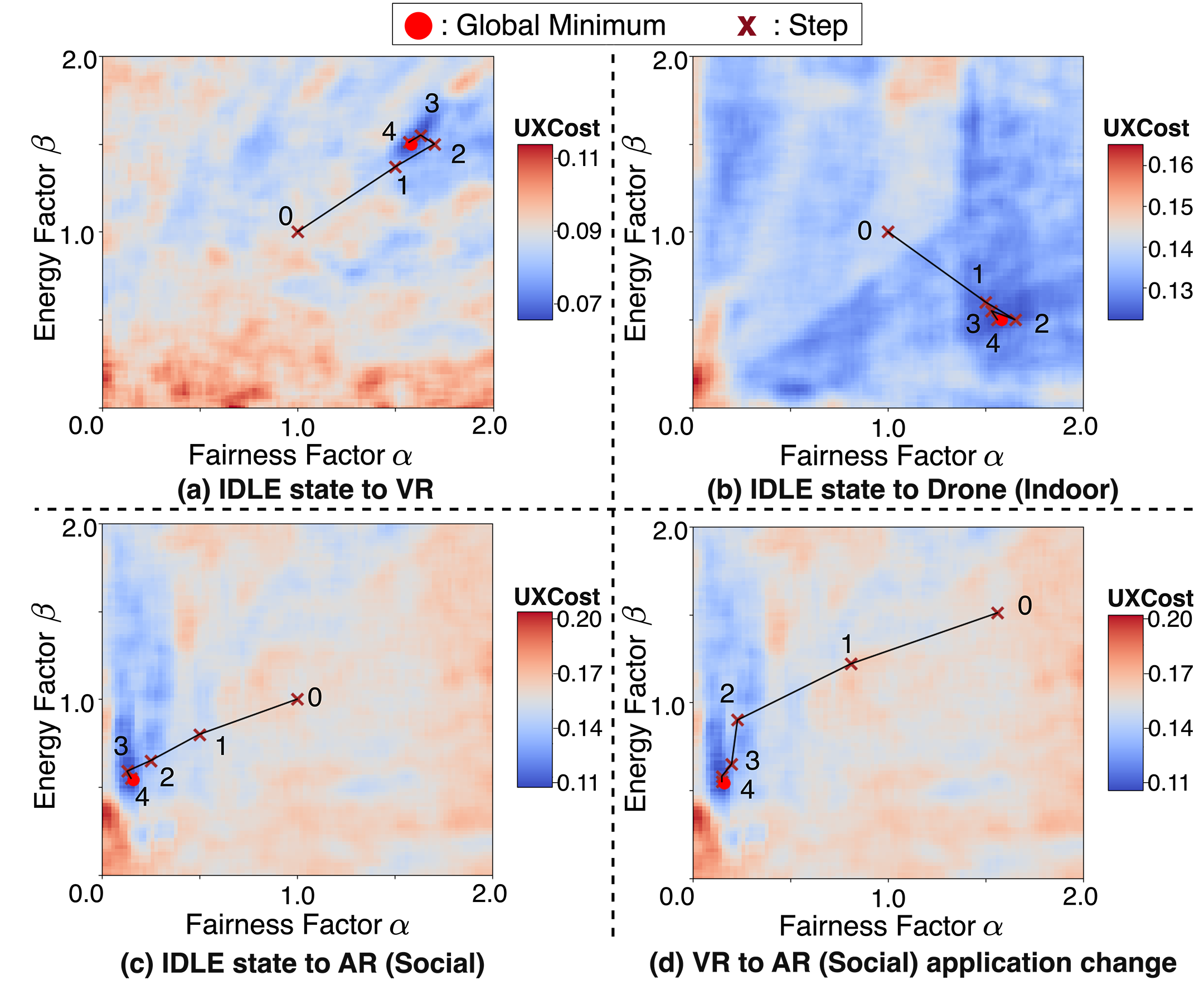}
    \caption{\scoreMetric parameter search. IDLE refers to the state after booting a system with random $\alpha$ and $\beta$ values. \vspace{0mm}}
    \label{fig:parameter_search}
\end{figure}

\betterparagraph{\scheduler finds near global optimum \scoreMetric parameters under workload changes}
~\autoref{fig:parameter_search} shows the \scoreMetric parameter (fairness and energy factors; $\alpha$ and $\beta$) search process on four workload change scenarios in the 4K 1OS+2WS setting. For the system booting to application cases (a, b, c), the parameters ($\alpha$ and $\beta$) are randomly initialized. Case (d) models a change in the runtime scenario from \vr to \ar, which sets the starting point from the locked parameters in (a).
On average across the workload change scenarios, \scheduler identified fairness and energy factor pairs that converge within 2\% of the global optimum in \costFunction space. The results show that our parameter optimization method successfully reaches a near-global optimum in various workload change cases. This is enabled by the well-conditioned search space with a constrained search range of the parameters within [0,2]. 

\begin{figure}[t]
    \centering
    \includegraphics[width=0.9\linewidth]{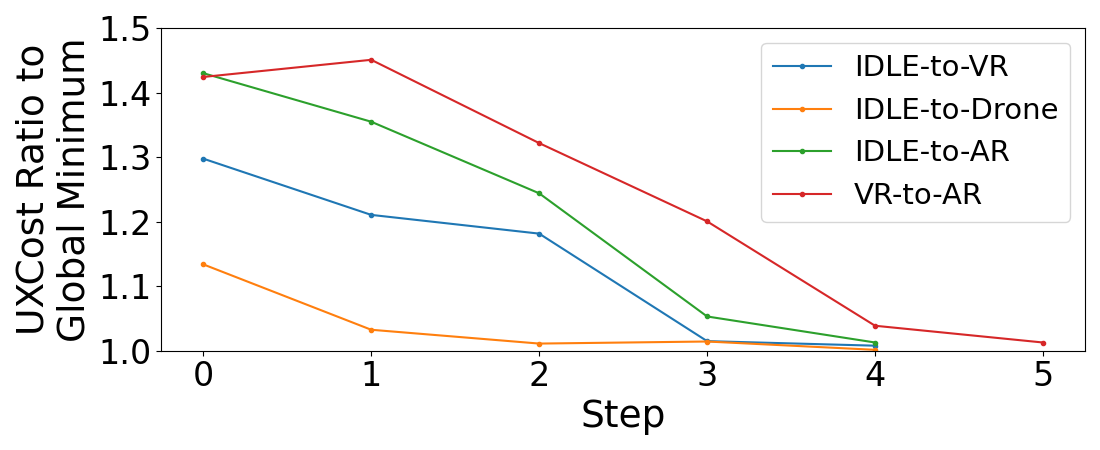}
    \vspace{-0.2cm}
    \caption{\scoreMetric parameter optimization converge \vspace{-2mm}}
    \label{fig:parameter_converge}
\end{figure}

\betterparagraph{\scheduler quickly adapts to workload changes}
~\autoref{fig:parameter_converge} shows the optimization process steps. 
For all cases, the optimization progress of \scoreMetric can improve more than 25\% of \costFunction in just two steps, and within five steps, the parameters of \scoreMetric converge within 2\% of the global minimum \costFunction.
This implies that a system with \scheduler is able to quickly adapt to workload changes while processing real-time workloads without significant overhead as the parameter optimization does not block the execution of workloads.

\begin{figure}[t]
    \centering
    \includegraphics[width=\linewidth]{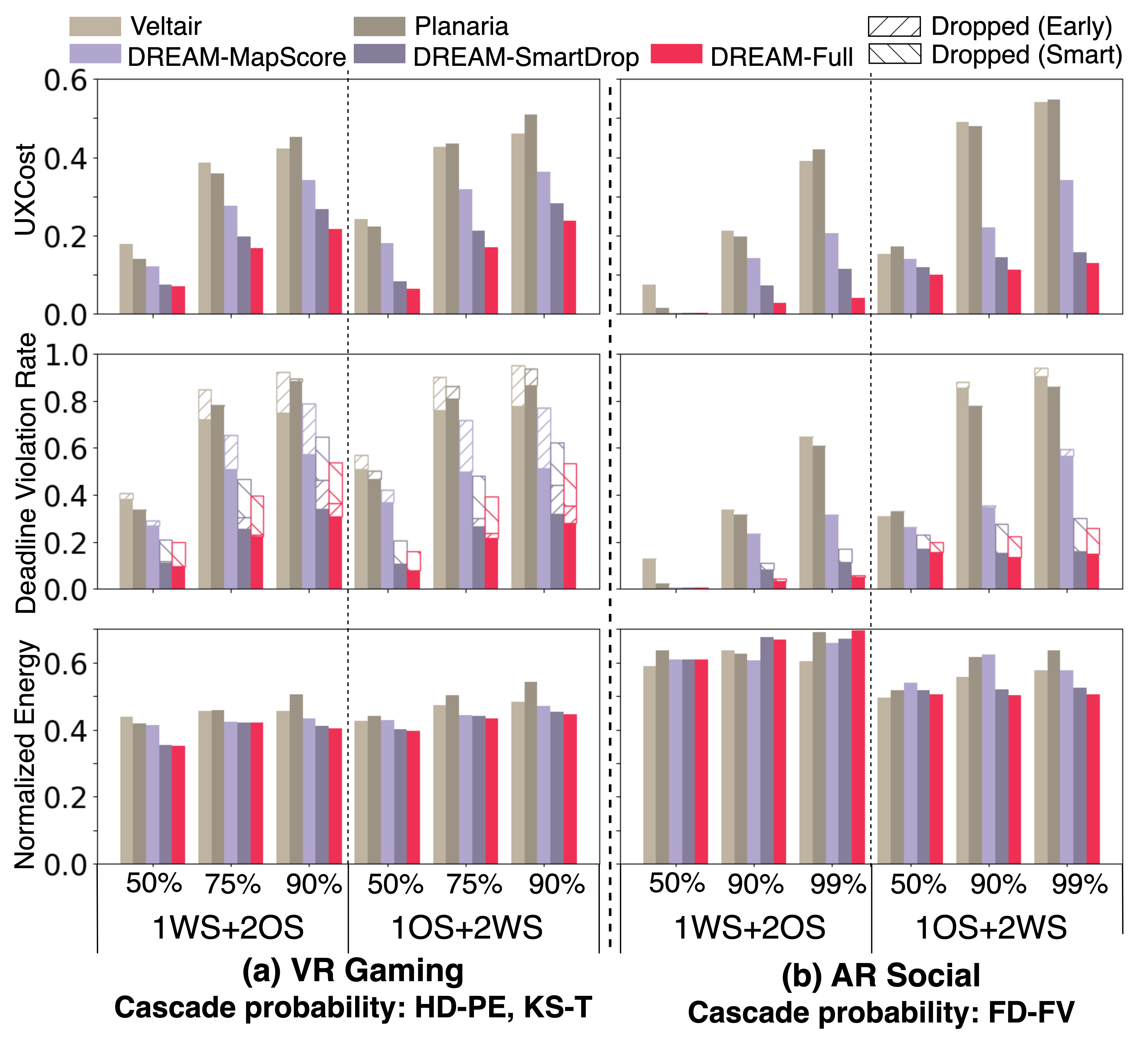}
    \caption{Performance comparison by differing ML cascade pipeline probability. \vspace{0mm}}
    \label{fig:cascade_prob}
\end{figure}

\betterparagraph{\scheduler is effective for dynamic workloads} 
To evaluate \scheduler and baselines in various dynamic workload scenarios, we vary the probability of the ML cascade pipeline 
of \vr and \ar from 50\% to 90\%, using 4K heterogeneous accelerators. 
As ~\autoref{fig:cascade_prob} shows, \scheduler consistently shows better performance than baselines.
The improvement is more significant under heavy system load.
In particular, \scheduler reduces \costFunction by 89.8\% compared to Veltair and 90.5\% compared to Planaria for \ar (99\%) running in the 1WS+2OS configuration, and by 77.1\% and 76.6\% for \ar (99\%) in 1OS+2WS.
We also observe that smart frame drop and Supernet switching are effective for dynamic workloads.
For instance, for \ar (99\%) running on the 1WS+2OS configuration, \smartdrop reduces \costFunction by 48.1\% over \dyscore, and \full further reduces it by 65.5\%. 
In the 1OS+2WS configuration, \smartdrop reduces \costFunction by 53.8\% over \dyscore, and \full shows a further reduction of 22.1\% compared to \smartdrop in 99\% probability, whereas 50\% probability shows 15.7\% and 15.3\%, respectively. These results support the efficacy of smart frame drop and Supernet switching for dynamic workloads.


\begin{figure}[t]
    \centering
    \includegraphics[width=1\linewidth]{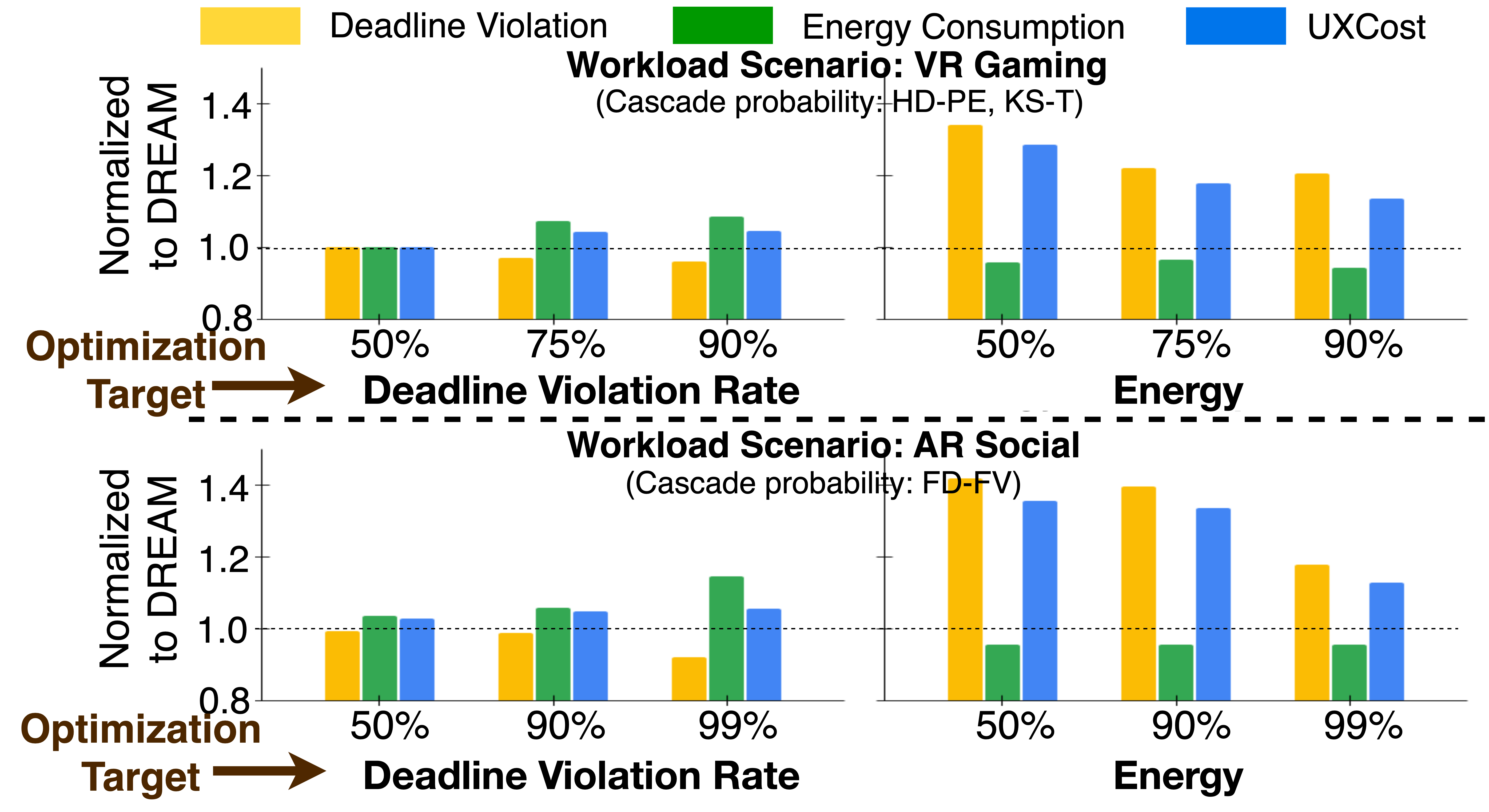}
    \caption{Deadline violation rate- and energy-only optimization results, normalized to \scheduler which uses \costFunction as an optimization metric. x-axis refers to the cascade probability.\vspace{-3mm}}
    \label{fig:uxcost_optim}
\end{figure}

\betterparagraph{\costFunction is an effective metric}
~\autoref{fig:uxcost_optim} shows \costFunction, deadline violation, and energy consumption rate result of using deadline violation rate or energy consumption rate as an optimization metric, compared to \scheduler which uses \costFunction for optimization. 
Using either one of the metrics may cause undesirable degradation in other metrics.
For \vr, optimizing using deadline violation rate can cause an 8.7\% increase in Energy consumption, which eventually leads to a 4.6\% increase in \costFunction. 
Optimizing Energy consumption would cause a 34.2\% increase in deadline violation rate for \vr with 50\% ML cascade pipeline probability, which eventually causes a 28.7\% increase in \costFunction.
For \ar, a 14.6\% increase in Energy consumption leads to a 5.7\% increase in \costFunction. Optimizing Energy consumption would cause a 41.9\% increase in the deadline violation rate for \ar with 50\% ML cascade pipeline probability, which eventually causes a 35.7\% increase in \costFunction.
However, \costFunction optimization helps balance the benefits between both metrics.

\begin{figure}[t]
    \centering
    \includegraphics[width=1\linewidth]{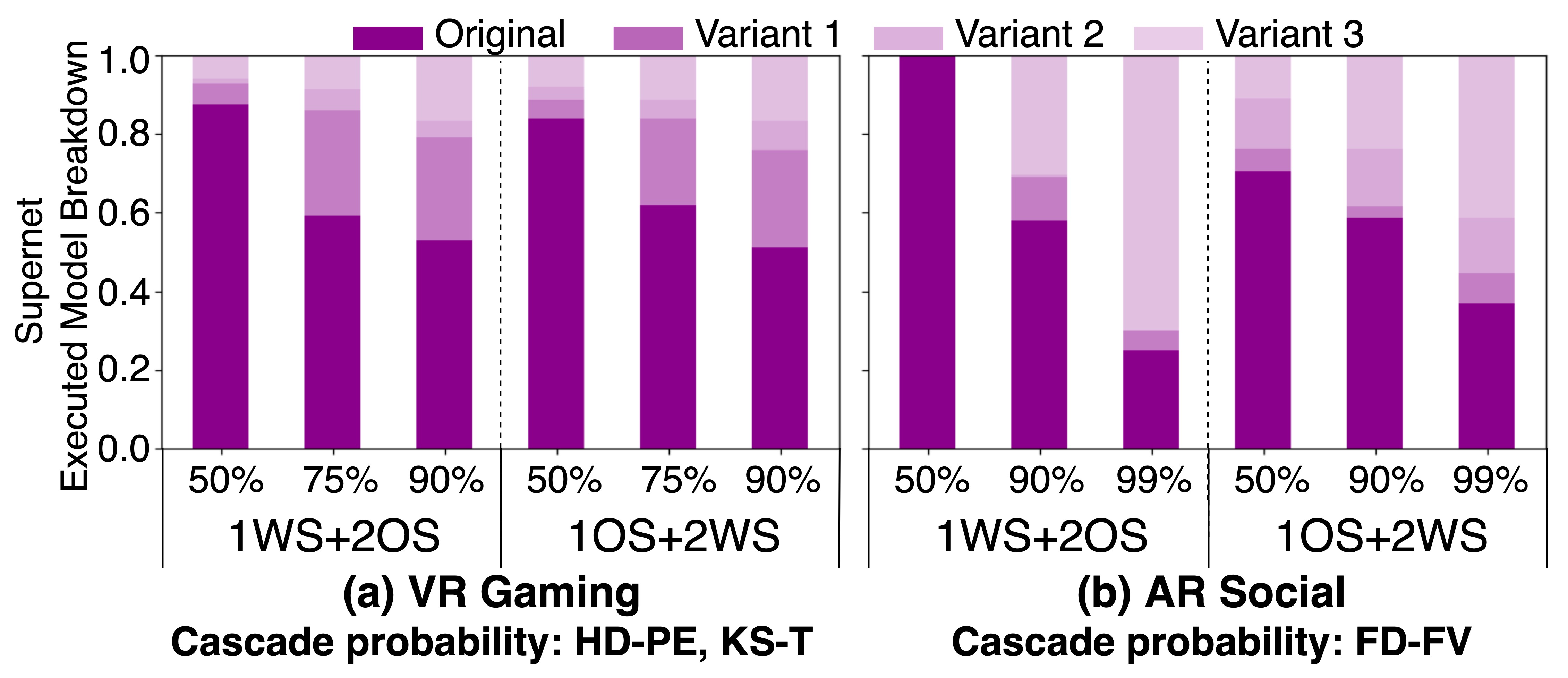}
    \caption{The executed Supernet subnetworks on heterogeneous 4K PE accelerators. \vspace{-4.5mm}}
    \label{fig:supernet_breakdown}
\end{figure}

\betterparagraph{Supernet switching is effective in reducing system load} 
~\autoref{fig:supernet_breakdown} shows the breakdown of subnets selected for a Supernet for context understanding. "Original" refers to the heaviest subnet (the default). Variants refer to lighter subnets deployed by \scheduler's Supernet switching algorithm. 
Under light system load (i.e., 50\% cascade probability), 
\scheduler mainly dispatches the original, more than 80\% for \vr, and 100\% for \ar on 1WS+2OS.
Under heavy system load, we observe that \scheduler actively utilizes Supernet switching. For example, more than 40\% of the executed Supernet models in \vr are smaller variants. For \ar, more than 60\% of Supernet models are lightweight variants.
These results show that \scheduler successfully detects system load and actively dispatches light-weighted variants to achieve global performance optimization, as ~\autoref{fig:uxcost} and~\autoref{fig:uxcost_homo} show.

\section{Related Works}
\label{label:related_works}
We summarize the related works in~\autoref{tab:prior_work} and discuss details of them in three categories as follows: 

\betterparagraph{Schedulers for multi-task DNN accelerators}
Prior dynamic schedulers on executing multiple workloads on DNN accelerator, such as Prema~\cite{PREMA}, propose time-multiplexing DNN execution.
However, such temporal co-location of workloads cannot utilize model parallelism for RTMM workloads and often involves preemption overheads.
On the other hand, other works proposed spatial co-location of multi-DNN workloads~\cite{kwon2021heterogeneous, magma, Veltair, Planaria, moca}.
Herald~\cite{kwon2021heterogeneous} and MAGMA~\cite{magma} proposed spatial partitioning of compute or memory resources of the DNN accelerator statically.
They both target maximizing the throughput of batched offline workload without latency target, thus not suitable for real-time scenarios with dynamicity.
Veltair~\cite{Veltair} employs a layer-blocking approach to avoid resource scheduling conflicts on general-purpose CPU clusters. 
Planaria~\cite{Planaria} proposes dynamic allocation of compute resources with a deadline-aware scheduler. 
\tobechecked{MoCA~\cite{moca} proposes dynamic memory resource partitioning with a deadline and compute-to-memory ratio aware scheduler.}
However, unlike \scheduler, those works do not consider energy and various workload dynamicity.

\begin{table}[t]
\centering
\caption{Comparison against prior real-time general-purpose scheduler and scheduler for DNN accelerator.\vspace{-1mm}}
\scalebox{0.67} {
\begin{tabular}{|c|l||c|c|c|}
\hline
\begin{tabular}[c]{@{}c@{}}\textbf{Target} \\ \textbf{Hardware}\end{tabular}
& \multicolumn{1}{c||}{\textbf{Works}}
& \begin{tabular}[c]{@{}c@{}}\textbf{Deadline}\\\textbf{Aware}\end{tabular}
& \begin{tabular}[c]{@{}c@{}}\textbf{Heterogeneity}\\\textbf{Aware}\end{tabular} 
& \begin{tabular}[c]{@{}c@{}}\textbf{Workload}\\\textbf{Adaptivity}\end{tabular}  \\
\hline
\hline
\multirow{4}{*}{
\begin{tabular}[c]{@{}c@{}}General\\Purpose\end{tabular}} 
    & \begin{tabular}[l]{@{}l@{}}Deadline-only \\ \cite{combined_task_message_scheduling, scheduling_exclusion_relation, allocation_complex_periodic_task, 4032340, 477248, 372795, spring_kernel}\end{tabular}
    & \cmark
    &
    & \\
\cline{2-5}
    & \begin{tabular}[l]{@{}l@{}}Harmony~\cite{harmony}, $HySARC^2$~\cite{resource_hetero_hybrid_scheduling},\\TTSA\cite{TTSA} \end{tabular}
    & \cmark
    & \cmark
    & \\
\cline{2-5}
    &Skip-over\cite{skip-over}, (n,m)~\cite{combining_nm_hard_scheduling}, (m,k)~\cite{m_k_firm_overload_management} 
    & \cmark                                                 
    &                               
    & \cmark \\
\cline{2-5}
    & Nexus~\cite{Nexus}, Clockwork~\cite{clockwork}
    & \cmark                                                       
    &                                                             
    & \cmark \\
\hline
\multirow{4}{*}{\begin{tabular}[c]{@{}c@{}}DNN\\Accelerator\end{tabular}}
    & MoCA~\cite{moca}, Veltair~\cite{Veltair}, Prema~\cite{PREMA}          
    & \cmark 
    &
    &
    \\
\cline{2-5}
    & Planaria~\cite{Planaria}
    & \cmark 
    & \cmark  
    & \\
\cline{2-5}
    & Herald~\cite{kwon2021heterogeneous}, MAGMA~\cite{magma}                
    &
    & \cmark 
    & \\
\cline{2-5}
    & \textbf{\scheduler (This work) }
    & \cmark
    & \cmark
    & \cmark  \\
\hline    
\end{tabular}
}
\label{tab:prior_work}
\vspace{-4mm}
\end{table}

\betterparagraph{Schedulers for general purpose hardware}
Many previous works in the operating system domain explored the scheduling problem on periodic tasks with constraints in real-time system~\cite{combined_task_message_scheduling, scheduling_exclusion_relation, allocation_complex_periodic_task, 4032340, 477248, 372795}.
However, they use offline static scheduling, which is not suitable for RTMM workloads with dynamicity.
%
Joint dynamic and static schedulers~\cite{368008, lehoczky1992optimal, ripoll1997optimal} have been proposed to reduce the overhead of dynamic scheduling while dealing with an aperiodic task.
However, such works target general-purpose systems and do not exploit the predictability of performances of ML workloads on accelerators.
Another work~\cite{resource_hetero_hybrid_scheduling} proposed algorithms with heterogeneity-aware workload clustering with a deadline-aware earliest-deadline-first algorithm. However, since workloads are pre-clustered before scheduling, this would make the scheduler suffer from resource conflict if workloads with similar computation demands are present. In addition, their scheduler does not consider the dynamic behavior of the workload.
Other cloud task schedulers ~\cite{hybrid_cloud_profit_maximization, TTSA} target different optimization goals than this work (e.g. maximizing economic profit of cloud).

\betterparagraph{Workload management for overloaded system}
Many works have proposed various frame drop techniques to enhance the overall system performance under heavy system load.
Skip-over~\cite{skip-over} employed a static rate-based frame skip technique, but such a static method does not consider model dependency, dynamicity, and other important aspects of RTMM workloads running on accelerators.

Some other works proposed priority-based frame drop methods~\cite{combining_nm_hard_scheduling, m_k_firm_overload_management}, which determines a subset of tasks to invoke based on off-line priority information. 
However, focusing on static information is not aligned with the RTMM workload dynamicity characteristics. Its high dynamicity can result in unnecessary or excessive frame drops with the possibility of high cost by forfeiting completed jobs for preceding models in a dependency chain. 
Some other works such as Nexus~\cite{Nexus}  and Clockwork~\cite{clockwork} target GPU clusters.
Nexus~\cite{Nexus} proposes to dynamically drop a subset of batches. However, a batch-focused approach is not tailored for RTMM workloads where each inference request mainly consists of a single batch.
Clockwork~\cite{clockwork} utilizes an FCFS model-wise scheduling method and drops requests upon arrival if the controller predicts that the request will not meet the deadline. However, the FCFS mechanism is not suitable for highly dynamic RTMM workloads, as discussed in our evaluation.


\section{Conclusion}
\label{sec:conclusion}

Emerging RTMM workloads introduce unique challenges to the ML system design, including real-time processing, complex model dependencies and heterogeneity, and various levels of dynamicity.
In this work, we clarified the types of dynamicity in RTMM workloads using a taxonomy in various granularities.
Based on the RTMM challenges and our insights on the dynamicity, we developed \scheduler that holistically considers all the challenges and types of dynamicity in RTMM workloads.
In our evaluation using industry-originated RTMM workloads, we observe the importance of such a holistic approach to RTMM workloads codified in \scheduler toward low deadline violation rate and energy consumption.
%
Also, we identify that considering global contexts of inference requests on different frames is another key factor, which helped \scheduler outperform other state-of-the-art dynamic schedulers~\cite{Planaria, Veltair} on RTMM workloads.
Finally, the benefits of Supernet switching we observed in our evaluation indicates that such ML algorithm-system software co-design methods can be future breakthroughs for other similar problems in ML system software.
%

%

\begin{acks}
We thank the anonymous reviewers for their insightful comments. We thank Ali Shafiee Ardestani and Krishnakumar Nair for constructive feedback. We thank Harshit Khaitan and Don Stark for their support.
\end{acks}


\bibliographystyle{ACM-Reference-Format}
\balance
\bibliography{ref}


\begin{thebibliography}{51}


\ifx \showCODEN    \undefined \def \showCODEN     #1{\unskip}     \fi
\ifx \showDOI      \undefined \def \showDOI       #1{#1}\fi
\ifx \showISBNx    \undefined \def \showISBNx     #1{\unskip}     \fi
\ifx \showISBNxiii \undefined \def \showISBNxiii  #1{\unskip}     \fi
\ifx \showISSN     \undefined \def \showISSN      #1{\unskip}     \fi
\ifx \showLCCN     \undefined \def \showLCCN      #1{\unskip}     \fi
\ifx \shownote     \undefined \def \shownote      #1{#1}          \fi
\ifx \showarticletitle \undefined \def \showarticletitle #1{#1}   \fi
\ifx \showURL      \undefined \def \showURL       {\relax}        \fi
\providecommand\bibfield[2]{#2}
\providecommand\bibinfo[2]{#2}
\providecommand\natexlab[1]{#1}
\providecommand\showeprint[2][]{arXiv:#2}

\bibitem[Abdelzaher and Shin(1999)]%
        {combined_task_message_scheduling}
\bibfield{author}{\bibinfo{person}{T.F. Abdelzaher} {and} \bibinfo{person}{K.G.
  Shin}.} \bibinfo{year}{1999}\natexlab{}.
\newblock \showarticletitle{Combined task and message scheduling in distributed
  real-time systems}.
\newblock \bibinfo{journal}{\emph{IEEE Transactions on Parallel and Distributed
  Systems}} (\bibinfo{year}{1999}).
\newblock


\bibitem[Bernat and Burns(1997)]%
        {combining_nm_hard_scheduling}
\bibfield{author}{\bibinfo{person}{G. Bernat} {and} \bibinfo{person}{A.
  Burns}.} \bibinfo{year}{1997}\natexlab{}.
\newblock \showarticletitle{Combining (/sub m//sup n/)-hard deadlines and dual
  priority scheduling}. In \bibinfo{booktitle}{\emph{Proceedings of the 18th
  IEEE Real-Time Systems Symposium}} \emph{(\bibinfo{series}{RTSS'97})}.
  \bibinfo{publisher}{IEEE}, \bibinfo{address}{San Francisco, CA, USA},
  \bibinfo{pages}{46--57}.
\newblock
\urldef\tempurl%
\url{https://doi.org/10.1109/REAL.1997.641268}
\showDOI{\tempurl}


\bibitem[Burchard et~al\mbox{.}(1995)]%
        {477248}
\bibfield{author}{\bibinfo{person}{A. Burchard}, \bibinfo{person}{J.
  Liebeherr}, \bibinfo{person}{Yingfeng Oh}, {and} \bibinfo{person}{S.H. Son}.}
  \bibinfo{year}{1995}\natexlab{}.
\newblock \showarticletitle{New strategies for assigning real-time tasks to
  multiprocessor systems}.
\newblock \bibinfo{journal}{\emph{IEEE Trans. Comput.}} (\bibinfo{year}{1995}).
\newblock


\bibitem[Cai et~al\mbox{.}(2020)]%
        {once-for-all}
\bibfield{author}{\bibinfo{person}{Han Cai}, \bibinfo{person}{Chuang Gan},
  \bibinfo{person}{Tianzhe Wang}, \bibinfo{person}{Zhekai Zhang}, {and}
  \bibinfo{person}{Song Han}.} \bibinfo{year}{2020}\natexlab{}.
\newblock \showarticletitle{Once for All: Train One Network and Specialize it
  for Efficient Deployment}. In \bibinfo{booktitle}{\emph{International
  Conference on Learning Representations}} \emph{(\bibinfo{series}{ICLR
  2020})}.
\newblock
\urldef\tempurl%
\url{https://openreview.net/pdf?id=HylxE1HKwS}
\showURL{%
\tempurl}


\bibitem[Cho et~al\mbox{.}(2006)]%
        {4032340}
\bibfield{author}{\bibinfo{person}{Hyeonjoong Cho}, \bibinfo{person}{Binoy
  Ravindran}, {and} \bibinfo{person}{E.~Douglas Jensen}.}
  \bibinfo{year}{2006}\natexlab{}.
\newblock \showarticletitle{An Optimal Real-Time Scheduling Algorithm for
  Multiprocessors}. In \bibinfo{booktitle}{\emph{IEEE International Real-Time
  Systems Symposium}} \emph{(\bibinfo{series}{RTSS 2006})}.
  \bibinfo{publisher}{IEEE}, \bibinfo{address}{Rio de Janeiro, Brazil},
  \bibinfo{pages}{101--110}.
\newblock
\urldef\tempurl%
\url{https://doi.org/10.1109/RTSS.2006.10}
\showDOI{\tempurl}


\bibitem[Choi and Rhu(2020)]%
        {PREMA}
\bibfield{author}{\bibinfo{person}{Yujeong Choi} {and} \bibinfo{person}{Minsoo
  Rhu}.} \bibinfo{year}{2020}\natexlab{}.
\newblock \showarticletitle{Prema: A predictive multi-task scheduling algorithm
  for preemptible neural processing units}. In
  \bibinfo{booktitle}{\emph{Proceedings of 2020 IEEE International Symposium on
  High Performance Computer Architecture}} \emph{(\bibinfo{series}{HPCA
  2020})}. \bibinfo{publisher}{IEEE}, \bibinfo{address}{San Diego, CA, USA},
  \bibinfo{pages}{220--233}.
\newblock
\urldef\tempurl%
\url{https://doi.org/10.1109/HPCA47549.2020.00027}
\showDOI{\tempurl}


\bibitem[Du et~al\mbox{.}(2015)]%
        {du2015shidiannao}
\bibfield{author}{\bibinfo{person}{Zidong Du}, \bibinfo{person}{Robert
  Fasthuber}, \bibinfo{person}{Tianshi Chen}, \bibinfo{person}{Paolo Ienne},
  \bibinfo{person}{Ling Li}, \bibinfo{person}{Tao Luo},
  \bibinfo{person}{Xiaobing Feng}, \bibinfo{person}{Yunji Chen}, {and}
  \bibinfo{person}{Olivier Temam}.} \bibinfo{year}{2015}\natexlab{}.
\newblock \showarticletitle{ShiDianNao: Shifting vision processing closer to
  the sensor}. In \bibinfo{booktitle}{\emph{Proceedings of the ACM/IEEE 42nd
  Annual International Symposium on Computer Architecture}}
  \emph{(\bibinfo{series}{ISCA 2015})}. \bibinfo{publisher}{IEEE},
  \bibinfo{address}{Portland, OR, USA}, \bibinfo{pages}{92--104}.
\newblock
\urldef\tempurl%
\url{https://doi.org/10.1145/2749469.2750389}
\showDOI{\tempurl}


\bibitem[Ghodrati et~al\mbox{.}(2020)]%
        {Planaria}
\bibfield{author}{\bibinfo{person}{Soroush Ghodrati},
  \bibinfo{person}{Byung~Hoon Ahn}, \bibinfo{person}{Joon~Kyung Kim},
  \bibinfo{person}{Sean Kinzer}, \bibinfo{person}{Brahmendra~Reddy Yatham},
  \bibinfo{person}{Navateja Alla}, \bibinfo{person}{Hardik Sharma},
  \bibinfo{person}{Mohammad Alian}, \bibinfo{person}{Eiman Ebrahimi},
  \bibinfo{person}{Nam~Sung Kim}, {et~al\mbox{.}}}
  \bibinfo{year}{2020}\natexlab{}.
\newblock \showarticletitle{Planaria: Dynamic architecture fission for spatial
  multi-tenant acceleration of deep neural networks}. In
  \bibinfo{booktitle}{\emph{Proceedings of The 53rd IEEE/ACM International
  Symposium on Microarchitecture}} \emph{(\bibinfo{series}{MICRO 2020})}.
  \bibinfo{publisher}{IEEE}, \bibinfo{address}{Athens, Greece},
  \bibinfo{pages}{681--697}.
\newblock
\urldef\tempurl%
\url{https://doi.org/10.1109/MICRO50266.2020.00062}
\showDOI{\tempurl}


\bibitem[Gujarati et~al\mbox{.}(2020)]%
        {clockwork}
\bibfield{author}{\bibinfo{person}{Arpan Gujarati}, \bibinfo{person}{Reza
  Karimi}, \bibinfo{person}{Safya Alzayat}, \bibinfo{person}{Wei Hao},
  \bibinfo{person}{Antoine Kaufmann}, \bibinfo{person}{Ymir Vigfusson}, {and}
  \bibinfo{person}{Jonathan Mace}.} \bibinfo{year}{2020}\natexlab{}.
\newblock \showarticletitle{Serving $\{$DNNs$\}$ like clockwork: Performance
  predictability from the bottom up}. In \bibinfo{booktitle}{\emph{Proceedings
  of the 14th USENIX Symposium on Operating Systems Design and Implementation}}
  \emph{(\bibinfo{series}{OSDI 20})}. \bibinfo{publisher}{USENIX Association},
  \bibinfo{address}{Banff, Canada}, \bibinfo{pages}{443--462}.
\newblock
\urldef\tempurl%
\url{https://doi.org/10.5555/3488766.3488791}
\showDOI{\tempurl}


\bibitem[He et~al\mbox{.}(2018)]%
        {focallengthdepth}
\bibfield{author}{\bibinfo{person}{Lei He}, \bibinfo{person}{Guanghui Wang},
  {and} \bibinfo{person}{Zhanyi Hu}.} \bibinfo{year}{2018}\natexlab{}.
\newblock \showarticletitle{Learning depth from single images with deep neural
  network embedding focal length}.
\newblock \bibinfo{journal}{\emph{IEEE Transactions on Image Processing}}
  \bibinfo{volume}{27}, \bibinfo{number}{9} (\bibinfo{year}{2018}),
  \bibinfo{pages}{4676--4689}.
\newblock
\urldef\tempurl%
\url{https://doi.org/10.1109/TIP.2018.2832296}
\showDOI{\tempurl}


\bibitem[Kao and Krishna(2022)]%
        {magma}
\bibfield{author}{\bibinfo{person}{Sheng-Chun Kao} {and}
  \bibinfo{person}{Tushar Krishna}.} \bibinfo{year}{2022}\natexlab{}.
\newblock \showarticletitle{Magma: An optimization framework for mapping
  multiple dnns on multiple accelerator cores}. In
  \bibinfo{booktitle}{\emph{Proceedings of the 2022 IEEE International
  Symposium on High-Performance Computer Architecture}}
  \emph{(\bibinfo{series}{HPCA 2022})}. IEEE, \bibinfo{publisher}{IEEE},
  \bibinfo{address}{Seoul, South Korea}, \bibinfo{pages}{814--830}.
\newblock
\urldef\tempurl%
\url{https://doi.org/10.1109/HPCA53966.2022.00065}
\showDOI{\tempurl}


\bibitem[Kim et~al\mbox{.}(2023)]%
        {moca}
\bibfield{author}{\bibinfo{person}{Seah Kim}, \bibinfo{person}{Hasan Genc},
  \bibinfo{person}{Vadim~Vadimovich Nikiforov}, \bibinfo{person}{Krste
  Asanovi{\'c}}, \bibinfo{person}{Borivoje Nikoli{\'c}}, {and}
  \bibinfo{person}{Yakun~Sophia Shao}.} \bibinfo{year}{2023}\natexlab{}.
\newblock \showarticletitle{MoCA: Memory-Centric, Adaptive Execution for
  Multi-Tenant Deep Neural Networks}. In \bibinfo{booktitle}{\emph{Proceedings
  of the 2023 IEEE International Symposium on High-Performance Computer
  Architecture}} \emph{(\bibinfo{series}{HPCA 2023})}.
  \bibinfo{publisher}{IEEE}, \bibinfo{address}{Montreal, Canada},
  \bibinfo{pages}{828--841}.
\newblock
\urldef\tempurl%
\url{https://doi.org/10.1109/HPCA56546.2023.10071035}
\showDOI{\tempurl}


\bibitem[Koren and Shasha(1995)]%
        {skip-over}
\bibfield{author}{\bibinfo{person}{Gilad Koren} {and} \bibinfo{person}{Dennis
  Shasha}.} \bibinfo{year}{1995}\natexlab{}.
\newblock \showarticletitle{Skip-over: Algorithms and complexity for overloaded
  systems that allow skips}. In \bibinfo{booktitle}{\emph{Proceedings of the
  16th IEEE Real-Time Systems Symposium}} \emph{(\bibinfo{series}{RTSS'95})}.
  \bibinfo{publisher}{IEEE}, \bibinfo{address}{Pisa, Italy},
  \bibinfo{pages}{110--117}.
\newblock
\urldef\tempurl%
\url{https://doi.org/10.1109/REAL.1995.495201}
\showDOI{\tempurl}


\bibitem[Kosta et~al\mbox{.}(2022)]%
        {RAPID-RL}
\bibfield{author}{\bibinfo{person}{Adarsh~Kumar Kosta},
  \bibinfo{person}{Malik~Aqeel Anwar}, \bibinfo{person}{Priyadarshini Panda},
  \bibinfo{person}{Arijit Raychowdhury}, {and} \bibinfo{person}{Kaushik Roy}.}
  \bibinfo{year}{2022}\natexlab{}.
\newblock \showarticletitle{RAPID-RL: A Reconfigurable Architecture with
  Preemptive-Exits for Efficient Deep-Reinforcement Learning}. In
  \bibinfo{booktitle}{\emph{Proceedings of the 2022 International Conference on
  Robotics and Automation}} \emph{(\bibinfo{series}{ICRA 2022})}.
  \bibinfo{publisher}{IEEE}, \bibinfo{address}{Philadelphia, PA, USA},
  \bibinfo{pages}{7492--7498}.
\newblock
\urldef\tempurl%
\url{https://doi.org/10.1109/ICRA46639.2022.9812320}
\showDOI{\tempurl}


\bibitem[Kwon et~al\mbox{.}(2019)]%
        {kwon2019understanding}
\bibfield{author}{\bibinfo{person}{Hyoukjun Kwon}, \bibinfo{person}{Prasanth
  Chatarasi}, \bibinfo{person}{Michael Pellauer}, \bibinfo{person}{Angshuman
  Parashar}, \bibinfo{person}{Vivek Sarkar}, {and} \bibinfo{person}{Tushar
  Krishna}.} \bibinfo{year}{2019}\natexlab{}.
\newblock \showarticletitle{Understanding reuse, performance, and hardware cost
  of dnn dataflow: A data-centric approach}. In
  \bibinfo{booktitle}{\emph{Proceedings of the 52nd Annual IEEE/ACM
  International Symposium on Microarchitecture}} \emph{(\bibinfo{series}{MICRO
  2019})}. \bibinfo{publisher}{IEEE}, \bibinfo{address}{Columbus, OH, USA},
  \bibinfo{pages}{754--768}.
\newblock
\urldef\tempurl%
\url{https://doi.org/10.1145/3352460.3358252}
\showDOI{\tempurl}


\bibitem[Kwon et~al\mbox{.}(2021)]%
        {kwon2021heterogeneous}
\bibfield{author}{\bibinfo{person}{Hyoukjun Kwon}, \bibinfo{person}{Liangzhen
  Lai}, \bibinfo{person}{Michael Pellauer}, \bibinfo{person}{Tushar Krishna},
  \bibinfo{person}{Yu-Hsin Chen}, {and} \bibinfo{person}{Vikas Chandra}.}
  \bibinfo{year}{2021}\natexlab{}.
\newblock \showarticletitle{Heterogeneous dataflow accelerators for multi-DNN
  workloads}. In \bibinfo{booktitle}{\emph{Proceedings of the 2021 IEEE
  International Symposium on High-Performance Computer Architecture}}
  \emph{(\bibinfo{series}{HPCA 2021})}. \bibinfo{publisher}{IEEE},
  \bibinfo{address}{Seoul, South Korea}, \bibinfo{pages}{71--83}.
\newblock
\urldef\tempurl%
\url{https://doi.org/10.1109/HPCA51647.2021.00016}
\showDOI{\tempurl}


\bibitem[Kwon et~al\mbox{.}(2023)]%
        {xrbench}
\bibfield{author}{\bibinfo{person}{Hyoukjun Kwon},
  \bibinfo{person}{Krishnakumar Nair}, \bibinfo{person}{Jamin Seo},
  \bibinfo{person}{Jason Yik}, \bibinfo{person}{Debabrata Mohapatra},
  \bibinfo{person}{Dongyuan Zhan}, \bibinfo{person}{Jinook Song},
  \bibinfo{person}{Peter Capak}, \bibinfo{person}{Peizhao Zhang},
  \bibinfo{person}{Peter Vajda}, \bibinfo{person}{Colby Banbury},
  \bibinfo{person}{Mark Mazumder}, \bibinfo{person}{Liangzhen Lai},
  \bibinfo{person}{Ashish Sirasao}, \bibinfo{person}{Tushar Krishna},
  \bibinfo{person}{Harshit Khaitan}, \bibinfo{person}{Vikas Chandra}, {and}
  \bibinfo{person}{Vijay~Janapa Reddi}.} \bibinfo{year}{2023}\natexlab{}.
\newblock \showarticletitle{XRBench: An Extended Reality (XR) Machine Learning
  Benchmark Suite for the Metaverse}. In \bibinfo{booktitle}{\emph{Proceedings
  of the 5th Machine Learning and Systems Conference (MLSys 2023)}}
  \emph{(\bibinfo{series}{MLSys 2023})}. \bibinfo{address}{Miami, FL, USA}.
\newblock
\urldef\tempurl%
\url{https://proceedings.mlsys.org/paper_files/paper/2023/hash/baf570e47e7f4e314a9ffb72c4a5459c-Abstract-mlsys2023.html}
\showURL{%
\tempurl}


\bibitem[Lea et~al\mbox{.}(2017)]%
        {tcn}
\bibfield{author}{\bibinfo{person}{Colin Lea}, \bibinfo{person}{Michael~D
  Flynn}, \bibinfo{person}{Rene Vidal}, \bibinfo{person}{Austin Reiter}, {and}
  \bibinfo{person}{Gregory~D Hager}.} \bibinfo{year}{2017}\natexlab{}.
\newblock \showarticletitle{Temporal convolutional networks for action
  segmentation and detection}. In \bibinfo{booktitle}{\emph{Proceedings of the
  IEEE Conference on Computer Vision and Pattern Recognition}}
  \emph{(\bibinfo{series}{CVPR 2017})}. \bibinfo{address}{Honolulu, HI, USA},
  \bibinfo{pages}{156--165}.
\newblock
\urldef\tempurl%
\url{https://openaccess.thecvf.com/content_cvpr_2017/html/Lea_Temporal_Convolutional_Networks_CVPR_2017_paper.html}
\showURL{%
\tempurl}


\bibitem[Lehoczky and Ramos-Thuel(1992)]%
        {lehoczky1992optimal}
\bibfield{author}{\bibinfo{person}{J.P. Lehoczky} {and} \bibinfo{person}{S.
  Ramos-Thuel}.} \bibinfo{year}{1992}\natexlab{}.
\newblock \showarticletitle{An optimal algorithm for scheduling soft-aperiodic
  tasks in fixed-priority preemptive systems}. In
  \bibinfo{booktitle}{\emph{[1992] Proceedings Real-Time Systems Symposium}}
  \emph{(\bibinfo{series}{RTSS'92})}. \bibinfo{publisher}{IEEE},
  \bibinfo{address}{Phoenix, AZ, USA}, \bibinfo{pages}{110--123}.
\newblock
\urldef\tempurl%
\url{https://doi.org/10.1109/REAL.1992.242671}
\showDOI{\tempurl}


\bibitem[Liu et~al\mbox{.}(2016)]%
        {SSD}
\bibfield{author}{\bibinfo{person}{Wei Liu}, \bibinfo{person}{Dragomir
  Anguelov}, \bibinfo{person}{Dumitru Erhan}, \bibinfo{person}{Christian
  Szegedy}, \bibinfo{person}{Scott Reed}, \bibinfo{person}{Cheng-Yang Fu},
  {and} \bibinfo{person}{Alexander~C Berg}.} \bibinfo{year}{2016}\natexlab{}.
\newblock \showarticletitle{SSD: Single shot multibox detector}. In
  \bibinfo{booktitle}{\emph{Proceedings of the 14th European Conference}}
  \emph{(\bibinfo{series}{ECCV 2016})}. \bibinfo{publisher}{Amsterdam, the
  Netherlands}, \bibinfo{address}{Portland, OR, USA}, \bibinfo{pages}{21--37}.
\newblock
\urldef\tempurl%
\url{https://doi.org/10.1007/978-3-319-46448-0_2}
\showDOI{\tempurl}


\bibitem[Liu et~al\mbox{.}(2022)]%
        {Veltair}
\bibfield{author}{\bibinfo{person}{Zihan Liu}, \bibinfo{person}{Jingwen Leng},
  \bibinfo{person}{Zhihui Zhang}, \bibinfo{person}{Quan Chen},
  \bibinfo{person}{Chao Li}, {and} \bibinfo{person}{Minyi Guo}.}
  \bibinfo{year}{2022}\natexlab{}.
\newblock \showarticletitle{VELTAIR: towards high-performance multi-tenant deep
  learning services via adaptive compilation and scheduling}. In
  \bibinfo{booktitle}{\emph{Proceedings of the 27th ACM International
  Conference on Architectural Support for Programming Languages and Operating
  Systems}} \emph{(\bibinfo{series}{ASPLOS 2022})}.
  \bibinfo{publisher}{Association for Computing Machinery (ACM)},
  \bibinfo{address}{Lausanne, Switzerland}, \bibinfo{pages}{388--401}.
\newblock
\urldef\tempurl%
\url{https://doi.org/10.1145/3503222.3507752}
\showDOI{\tempurl}


\bibitem[Madadi et~al\mbox{.}(2022)]%
        {handposenet}
\bibfield{author}{\bibinfo{person}{Meysam Madadi}, \bibinfo{person}{Sergio
  Escalera}, \bibinfo{person}{Xavier Bar{\'o}}, {and} \bibinfo{person}{Jordi
  Gonz{\`a}lez}.} \bibinfo{year}{2022}\natexlab{}.
\newblock \showarticletitle{End-to-end global to local convolutional neural
  network learning for hand pose recovery in depth data}.
\newblock \bibinfo{journal}{\emph{IET Computer Vision}} \bibinfo{volume}{16},
  \bibinfo{number}{1} (\bibinfo{year}{2022}), \bibinfo{pages}{50--66}.
\newblock
\urldef\tempurl%
\url{https://doi.org/10.1049/cvi2.12064}
\showDOI{\tempurl}


\bibitem[Nagrani et~al\mbox{.}(2017)]%
        {voxceleb}
\bibfield{author}{\bibinfo{person}{Arsha Nagrani}, \bibinfo{person}{Joon~Son
  Chung}, {and} \bibinfo{person}{Andrew Zisserman}.}
  \bibinfo{year}{2017}\natexlab{}.
\newblock \showarticletitle{{VoxCeleb: A Large-Scale Speaker Identification
  Dataset}}. In \bibinfo{booktitle}{\emph{Procedings of the Interspeech 2017}}
  \emph{(\bibinfo{series}{Interspeech 2017})}. \bibinfo{address}{Stockholm,
  Sweden}, \bibinfo{pages}{2616--2620}.
\newblock
\urldef\tempurl%
\url{https://doi.org/10.21437/Interspeech.2017-950}
\showDOI{\tempurl}


\bibitem[NVIDIA(2017)]%
        {nvdla}
\bibfield{author}{\bibinfo{person}{NVIDIA}.} \bibinfo{year}{2017}\natexlab{}.
\newblock \bibinfo{title}{NVDLA Deep Learning Accelerator}.
\newblock \bibinfo{howpublished}{Retrived from \url{http://nvdla.org}}.
\newblock


\bibitem[Parashar et~al\mbox{.}(2019)]%
        {parashar2019timeloop}
\bibfield{author}{\bibinfo{person}{Angshuman Parashar},
  \bibinfo{person}{Priyanka Raina}, \bibinfo{person}{Yakun~Sophia Shao},
  \bibinfo{person}{Yu-Hsin Chen}, \bibinfo{person}{Victor~A Ying},
  \bibinfo{person}{Anurag Mukkara}, \bibinfo{person}{Rangharajan Venkatesan},
  \bibinfo{person}{Brucek Khailany}, \bibinfo{person}{Stephen~W Keckler}, {and}
  \bibinfo{person}{Joel Emer}.} \bibinfo{year}{2019}\natexlab{}.
\newblock \showarticletitle{Timeloop: A systematic approach to dnn accelerator
  evaluation}. In \bibinfo{booktitle}{\emph{2019 IEEE international symposium
  on performance analysis of systems and software}}
  \emph{(\bibinfo{series}{ISPASS 2019})}. \bibinfo{publisher}{IEEE},
  \bibinfo{address}{Madison, WI, USA}, \bibinfo{pages}{304--315}.
\newblock
\urldef\tempurl%
\url{https://doi.org/10.1109/ISPASS.2019.00042}
\showDOI{\tempurl}


\bibitem[Ramamritham(1990)]%
        {allocation_complex_periodic_task}
\bibfield{author}{\bibinfo{person}{K. Ramamritham}.}
  \bibinfo{year}{1990}\natexlab{}.
\newblock \showarticletitle{Allocation and scheduling of complex periodic
  tasks}. In \bibinfo{booktitle}{\emph{Proceedings of the 10th International
  Conference on Distributed Computing Systems}} \emph{(\bibinfo{series}{ICDCS
  1990})}. \bibinfo{publisher}{IEEE}, \bibinfo{address}{Los Alamitos, CA, USA},
  \bibinfo{pages}{108--109}.
\newblock
\urldef\tempurl%
\url{https://doi.org/10.1109/ICDCS.1990.89256}
\showDOI{\tempurl}


\bibitem[Ramamritham(1995)]%
        {372795}
\bibfield{author}{\bibinfo{person}{Krithi Ramamritham}.}
  \bibinfo{year}{1995}\natexlab{}.
\newblock \showarticletitle{Allocation and scheduling of precedence-related
  periodic tasks}.
\newblock \bibinfo{journal}{\emph{IEEE Transactions on Parallel and Distributed
  Systems}} \bibinfo{volume}{6}, \bibinfo{number}{4} (\bibinfo{year}{1995}),
  \bibinfo{pages}{412--420}.
\newblock
\urldef\tempurl%
\url{https://doi.org/10.1109/71.372795}
\showDOI{\tempurl}


\bibitem[Ramanathan(1999)]%
        {m_k_firm_overload_management}
\bibfield{author}{\bibinfo{person}{Parameswaran Ramanathan}.}
  \bibinfo{year}{1999}\natexlab{}.
\newblock \showarticletitle{Overload management in real-time control
  applications using (m, k)-firm guarantee}.
\newblock \bibinfo{journal}{\emph{IEEE Transactions on parallel and distributed
  systems}} \bibinfo{volume}{10}, \bibinfo{number}{6} (\bibinfo{year}{1999}),
  \bibinfo{pages}{549--559}.
\newblock
\urldef\tempurl%
\url{https://doi.org/10.1109/71.774906}
\showDOI{\tempurl}


\bibitem[Reddi et~al\mbox{.}(2020)]%
        {MLperf}
\bibfield{author}{\bibinfo{person}{Vijay~Janapa Reddi},
  \bibinfo{person}{Christine Cheng}, \bibinfo{person}{David Kanter},
  \bibinfo{person}{Peter Mattson}, \bibinfo{person}{Guenther Schmuelling},
  \bibinfo{person}{Carole-Jean Wu}, \bibinfo{person}{Brian Anderson},
  \bibinfo{person}{Maximilien Breughe}, \bibinfo{person}{Mark Charlebois},
  \bibinfo{person}{William Chou}, \bibinfo{person}{Ramesh Chukka},
  \bibinfo{person}{Cody Coleman}, \bibinfo{person}{Sam Davis},
  \bibinfo{person}{Pan Deng}, \bibinfo{person}{Greg Diamos},
  \bibinfo{person}{Jared Duke}, \bibinfo{person}{Dave Fick},
  \bibinfo{person}{J.~Scott Gardner}, \bibinfo{person}{Itay Hubara},
  \bibinfo{person}{Sachin Idgunji}, \bibinfo{person}{Thomas~B. Jablin},
  \bibinfo{person}{Jeff Jiao}, \bibinfo{person}{Tom~St. John},
  \bibinfo{person}{Pankaj Kanwar}, \bibinfo{person}{David Lee},
  \bibinfo{person}{Jeffery Liao}, \bibinfo{person}{Anton Lokhmotov},
  \bibinfo{person}{Francisco Massa}, \bibinfo{person}{Peng Meng},
  \bibinfo{person}{Paulius Micikevicius}, \bibinfo{person}{Colin Osborne},
  \bibinfo{person}{Gennady Pekhimenko}, \bibinfo{person}{Arun Tejusve~Raghunath
  Rajan}, \bibinfo{person}{Dilip Sequeira}, \bibinfo{person}{Ashish Sirasao},
  \bibinfo{person}{Fei Sun}, \bibinfo{person}{Hanlin Tang},
  \bibinfo{person}{Michael Thomson}, \bibinfo{person}{Frank Wei},
  \bibinfo{person}{Ephrem Wu}, \bibinfo{person}{Lingjie Xu},
  \bibinfo{person}{Koichi Yamada}, \bibinfo{person}{Bing Yu},
  \bibinfo{person}{George Yuan}, \bibinfo{person}{Aaron Zhong},
  \bibinfo{person}{Peizhao Zhang}, {and} \bibinfo{person}{Yuchen Zhou}.}
  \bibinfo{year}{2020}\natexlab{}.
\newblock \showarticletitle{Mlperf inference benchmark}. In
  \bibinfo{booktitle}{\emph{Proceddings of the 47th ACM/IEEE Annual
  International Symposium on Computer Architecture}}
  \emph{(\bibinfo{series}{ISCA 2020})}. IEEE, \bibinfo{publisher}{IEEE},
  \bibinfo{address}{Valencia, Spain}, \bibinfo{pages}{446--459}.
\newblock
\urldef\tempurl%
\url{https://doi.org/10.1109/ISCA45697.2020.00045}
\showDOI{\tempurl}


\bibitem[Ripoll et~al\mbox{.}(1997)]%
        {ripoll1997optimal}
\bibfield{author}{\bibinfo{person}{Ismael Ripoll}, \bibinfo{person}{Alfons
  Crespo}, {and} \bibinfo{person}{Ana Garcia-Fornes}.}
  \bibinfo{year}{1997}\natexlab{}.
\newblock \showarticletitle{An optimal algorithm for scheduling soft aperiodic
  tasks in dynamic-priority preemptive systems}.
\newblock \bibinfo{journal}{\emph{IEEE Transactions on Software Engineering}}
  \bibinfo{volume}{23}, \bibinfo{number}{6} (\bibinfo{year}{1997}),
  \bibinfo{pages}{388--400}.
\newblock
\urldef\tempurl%
\url{https://doi.org/10.1109/32.601081}
\showDOI{\tempurl}


\bibitem[Shen et~al\mbox{.}(2019)]%
        {Nexus}
\bibfield{author}{\bibinfo{person}{Haichen Shen}, \bibinfo{person}{Lequn Chen},
  \bibinfo{person}{Yuchen Jin}, \bibinfo{person}{Liangyu Zhao},
  \bibinfo{person}{Bingyu Kong}, \bibinfo{person}{Matthai Philipose},
  \bibinfo{person}{Arvind Krishnamurthy}, {and} \bibinfo{person}{Ravi
  Sundaram}.} \bibinfo{year}{2019}\natexlab{}.
\newblock \showarticletitle{Nexus: A GPU Cluster Engine for Accelerating
  DNN-Based Video Analysis}. In \bibinfo{booktitle}{\emph{Proceedings of the
  27th ACM Symposium on Operating Systems Principles}}
  \emph{(\bibinfo{series}{SOSP'19})}. \bibinfo{publisher}{ACM},
  \bibinfo{address}{Huntsville, Ontario, Canada}, \bibinfo{pages}{322--337}.
\newblock
\urldef\tempurl%
\url{https://doi.org/10.1145/3341301.3359658}
\showDOI{\tempurl}


\bibitem[Smolyanskiy et~al\mbox{.}(2017)]%
        {trailnet}
\bibfield{author}{\bibinfo{person}{Nikolai Smolyanskiy},
  \bibinfo{person}{Alexey Kamenev}, \bibinfo{person}{Jeffrey Smith}, {and}
  \bibinfo{person}{Stan Birchfield}.} \bibinfo{year}{2017}\natexlab{}.
\newblock \showarticletitle{Toward low-flying autonomous MAV trail navigation
  using deep neural networks for environmental awareness}. In
  \bibinfo{booktitle}{\emph{Proceedings of the 2017 IEEE/RSJ International
  Conference on Intelligent Robots and Systems}} \emph{(\bibinfo{series}{IROS
  2017})}. IEEE, \bibinfo{publisher}{IEEE}, \bibinfo{address}{Vancouver, BC,
  Canada}, \bibinfo{pages}{4241--4247}.
\newblock
\urldef\tempurl%
\url{https://doi.org/10.1109/IROS.2017.8206285}
\showDOI{\tempurl}


\bibitem[Stankovic and Ramamritham(1989)]%
        {spring_kernel}
\bibfield{author}{\bibinfo{person}{John~A. Stankovic} {and}
  \bibinfo{person}{Krithi Ramamritham}.} \bibinfo{year}{1989}\natexlab{}.
\newblock \showarticletitle{The Spring kernel: A new paradigm for real-time
  operating systems}.
\newblock \bibinfo{journal}{\emph{ACM SIGOPS Operating Systems Review}}
  \bibinfo{volume}{23}, \bibinfo{number}{3} (\bibinfo{year}{1989}),
  \bibinfo{pages}{54--71}.
\newblock
\showISSN{0163-5980}
\urldef\tempurl%
\url{https://doi.org/10.1145/71021.71024}
\showDOI{\tempurl}


\bibitem[Strosnider et~al\mbox{.}(1995)]%
        {368008}
\bibfield{author}{\bibinfo{person}{Jay~K. Strosnider}, \bibinfo{person}{John~P.
  Lehoczky}, {and} \bibinfo{person}{Lui Sha}.} \bibinfo{year}{1995}\natexlab{}.
\newblock \showarticletitle{The deferrable server algorithm for enhanced
  aperiodic responsiveness in hard real-time environments}.
\newblock \bibinfo{journal}{\emph{IEEE Trans. Comput.}} \bibinfo{volume}{44},
  \bibinfo{number}{1} (\bibinfo{date}{Jan.} \bibinfo{year}{1995}),
  \bibinfo{pages}{73--91}.
\newblock
\urldef\tempurl%
\url{https://doi.org/10.1109/12.368008}
\showDOI{\tempurl}


\bibitem[Tang and Lin(2018)]%
        {kws-res26}
\bibfield{author}{\bibinfo{person}{Raphael Tang} {and} \bibinfo{person}{Jimmy
  Lin}.} \bibinfo{year}{2018}\natexlab{}.
\newblock \showarticletitle{Deep residual learning for small-footprint keyword
  spotting}. In \bibinfo{booktitle}{\emph{Proceedings of the 2018 IEEE
  International Conference on Acoustics, Speech and Signal Processing}}
  \emph{(\bibinfo{series}{ICASSP 2018})}. IEEE, \bibinfo{publisher}{IEEE},
  \bibinfo{address}{Calgary, AB, Canada}, \bibinfo{pages}{5484--5488}.
\newblock
\urldef\tempurl%
\url{https://doi.org/10.1109/ICASSP.2018.8462688}
\showDOI{\tempurl}


\bibitem[Teerapittayanon et~al\mbox{.}(2016)]%
        {branchynet}
\bibfield{author}{\bibinfo{person}{Surat Teerapittayanon},
  \bibinfo{person}{Bradley McDanel}, {and} \bibinfo{person}{Hsiang-Tsung
  Kung}.} \bibinfo{year}{2016}\natexlab{}.
\newblock \showarticletitle{Branchynet: Fast inference via early exiting from
  deep neural networks}. In \bibinfo{booktitle}{\emph{Proceedings of the 23rd
  international conference on pattern recognition}}
  \emph{(\bibinfo{series}{ICPR 2016})}. \bibinfo{publisher}{IEEE},
  \bibinfo{address}{Cancun, Mexico}, \bibinfo{pages}{2464--2469}.
\newblock
\urldef\tempurl%
\url{https://doi.org/10.1109/ICPR.2016.7900006}
\showDOI{\tempurl}


\bibitem[Tian et~al\mbox{.}(2019)]%
        {sosnet}
\bibfield{author}{\bibinfo{person}{Yurun Tian}, \bibinfo{person}{Xin Yu},
  \bibinfo{person}{Bin Fan}, \bibinfo{person}{Fuchao Wu}, \bibinfo{person}{Huub
  Heijnen}, {and} \bibinfo{person}{Vassileios Balntas}.}
  \bibinfo{year}{2019}\natexlab{}.
\newblock \showarticletitle{Sosnet: Second order similarity regularization for
  local descriptor learning}. In \bibinfo{booktitle}{\emph{Proceedings of the
  IEEE/CVF Conference on Computer Vision and Pattern Recognition}}
  \emph{(\bibinfo{series}{CVPR 2019})}. \bibinfo{publisher}{IEEE},
  \bibinfo{address}{Long Beach, CA, USA}, \bibinfo{pages}{11016--11025}.
\newblock
\urldef\tempurl%
\url{https://doi.org/10.1109/CVPR.2019.01127}
\showDOI{\tempurl}


\bibitem[Tong et~al\mbox{.}(2022)]%
        {tong2022enabling}
\bibfield{author}{\bibinfo{person}{Jianming Tong}, \bibinfo{person}{Yangyu
  Chen}, \bibinfo{person}{Yue Pan}, \bibinfo{person}{Abhimanyu Bambhaniya},
  \bibinfo{person}{Alind Khare}, \bibinfo{person}{Taekyung Heo},
  \bibinfo{person}{Alexey Tumanov}, {and} \bibinfo{person}{Tushar Krishna}.}
  \bibinfo{year}{2022}\natexlab{}.
\newblock \showarticletitle{Enabling Real-time DNN Switching via
  Weight-Sharing}. In \bibinfo{booktitle}{\emph{The 2nd Architecture, Compiler,
  and System Support for Multi-model DNN Workloads Workshop (ACSMD 2022)}}.
  \bibinfo{address}{New York, NY, USA}.
\newblock
\urldef\tempurl%
\url{https://research.facebook.com/file/703126461319360/enabling-real-time-dnn-switching-via-weight-sharing.pdf}
\showURL{%
\tempurl}


\bibitem[Vasile et~al\mbox{.}(2015)]%
        {resource_hetero_hybrid_scheduling}
\bibfield{author}{\bibinfo{person}{Mihaela-Andreea Vasile},
  \bibinfo{person}{Florin Pop}, \bibinfo{person}{Radu-Ioan Tutueanu},
  \bibinfo{person}{Valentin Cristea}, {and} \bibinfo{person}{Joanna
  Ko{\l}odziej}.} \bibinfo{year}{2015}\natexlab{}.
\newblock \showarticletitle{Resource-aware hybrid scheduling algorithm in
  heterogeneous distributed computing}.
\newblock \bibinfo{journal}{\emph{Future Generation Computer Systems}}
  \bibinfo{volume}{51} (\bibinfo{date}{Oct.} \bibinfo{year}{2015}),
  \bibinfo{pages}{61--71}.
\newblock
\showISSN{0167-739X}
\urldef\tempurl%
\url{https://doi.org/10.1016/j.future.2014.11.019}
\showDOI{\tempurl}


\bibitem[Veit and Belongie(2018)]%
        {conv-aig}
\bibfield{author}{\bibinfo{person}{Andreas Veit} {and} \bibinfo{person}{Serge
  Belongie}.} \bibinfo{year}{2018}\natexlab{}.
\newblock \showarticletitle{Convolutional networks with adaptive inference
  graphs}. In \bibinfo{booktitle}{\emph{Proceedings of the European Conference
  on Computer Vision}} \emph{(\bibinfo{series}{ECCV 2018})}.
  \bibinfo{publisher}{IEEE}, \bibinfo{address}{Munich, Germany},
  \bibinfo{pages}{3--18}.
\newblock
\urldef\tempurl%
\url{https://doi.org/10.1007/978-3-030-01246-5_1}
\showDOI{\tempurl}


\bibitem[Wang et~al\mbox{.}(2021)]%
        {alphanet}
\bibfield{author}{\bibinfo{person}{Dilin Wang}, \bibinfo{person}{Chengyue
  Gong}, \bibinfo{person}{Meng Li}, \bibinfo{person}{Qiang Liu}, {and}
  \bibinfo{person}{Vikas Chandra}.} \bibinfo{year}{2021}\natexlab{}.
\newblock \showarticletitle{AlphaNet: Improved Training of Supernets with
  Alpha-Divergence}. In \bibinfo{booktitle}{\emph{Proceedings of the 38th
  International Conference on Machine Learning}} \emph{(\bibinfo{series}{ICML
  2021})}. \bibinfo{publisher}{PMLR}, \bibinfo{pages}{10760--10771}.
\newblock
\urldef\tempurl%
\url{https://proceedings.mlr.press/v139/wang21i.html}
\showURL{%
\tempurl}


\bibitem[Wang et~al\mbox{.}(2018)]%
        {SkipNet}
\bibfield{author}{\bibinfo{person}{Xin Wang}, \bibinfo{person}{Fisher Yu},
  \bibinfo{person}{Zi-Yi Dou}, \bibinfo{person}{Trevor Darrell}, {and}
  \bibinfo{person}{Joseph~E Gonzalez}.} \bibinfo{year}{2018}\natexlab{}.
\newblock \showarticletitle{Skipnet: Learning dynamic routing in convolutional
  networks}. In \bibinfo{booktitle}{\emph{Proceedings of the European
  Conference on Computer Vision}} \emph{(\bibinfo{series}{ECCV 2018})}.
  \bibinfo{publisher}{Springer}, \bibinfo{address}{Munich, Germany},
  \bibinfo{pages}{409--424}.
\newblock
\urldef\tempurl%
\url{https://doi.org/10.1007/978-3-030-01261-8_25}
\showDOI{\tempurl}


\bibitem[Wu et~al\mbox{.}(2019)]%
        {fbnet}
\bibfield{author}{\bibinfo{person}{Bichen Wu}, \bibinfo{person}{Xiaoliang Dai},
  \bibinfo{person}{Peizhao Zhang}, \bibinfo{person}{Yanghan Wang},
  \bibinfo{person}{Fei Sun}, \bibinfo{person}{Yiming Wu},
  \bibinfo{person}{Yuandong Tian}, \bibinfo{person}{Peter Vajda},
  \bibinfo{person}{Yangqing Jia}, {and} \bibinfo{person}{Kurt Keutzer}.}
  \bibinfo{year}{2019}\natexlab{}.
\newblock \showarticletitle{Fbnet: Hardware-aware efficient convnet design via
  differentiable neural architecture search}. In
  \bibinfo{booktitle}{\emph{Proceedings of the IEEE/CVF conference on computer
  vision and pattern recognition}} \emph{(\bibinfo{series}{CVPR 2019})}.
  \bibinfo{publisher}{IEEE}, \bibinfo{address}{Long Beach, CA, USA},
  \bibinfo{pages}{10734--10742}.
\newblock
\urldef\tempurl%
\url{https://doi.org/10.1109/CVPR.2019.01099}
\showDOI{\tempurl}


\bibitem[Wu et~al\mbox{.}(2016)]%
        {gnmt}
\bibfield{author}{\bibinfo{person}{Yonghui Wu}, \bibinfo{person}{Mike
  Schuster}, \bibinfo{person}{Zhifeng Chen}, \bibinfo{person}{Quoc~V. Le},
  \bibinfo{person}{Mohammad Norouzi}, \bibinfo{person}{Wolfgang Macherey},
  \bibinfo{person}{Maxim Krikun}, \bibinfo{person}{Yuan Cao},
  \bibinfo{person}{Qin Gao}, \bibinfo{person}{Klaus Macherey},
  \bibinfo{person}{Jeff Klingner}, \bibinfo{person}{Apurva Shah},
  \bibinfo{person}{Melvin Johnson}, \bibinfo{person}{Xiaobing Liu},
  \bibinfo{person}{Lukasz Kaiser}, \bibinfo{person}{Stephan Gouws},
  \bibinfo{person}{Yoshikiyo Kato}, \bibinfo{person}{Taku Kudo},
  \bibinfo{person}{Hideto Kazawa}, \bibinfo{person}{Keith Stevens},
  \bibinfo{person}{George Kurian}, \bibinfo{person}{Nishant Patil},
  \bibinfo{person}{Wei Wang}, \bibinfo{person}{Cliff Young},
  \bibinfo{person}{Jason Smith}, \bibinfo{person}{Jason Riesa},
  \bibinfo{person}{Alex Rudnick}, \bibinfo{person}{Oriol Vinyals},
  \bibinfo{person}{Greg Corrado}, \bibinfo{person}{Macduff Hughes}, {and}
  \bibinfo{person}{Jeffrey Dean}.} \bibinfo{year}{2016}\natexlab{}.
\newblock \showarticletitle{Google's neural machine translation system:
  Bridging the gap between human and machine translation}.
\newblock  (\bibinfo{year}{2016}).
\newblock
\urldef\tempurl%
\url{https://doi.org/10.48550/arXiv.1609.08144}
\showDOI{\tempurl}
\showeprint{arXiv:1609.08144}


\bibitem[Wu et~al\mbox{.}(2018)]%
        {blockdrop}
\bibfield{author}{\bibinfo{person}{Zuxuan Wu}, \bibinfo{person}{Tushar
  Nagarajan}, \bibinfo{person}{Abhishek Kumar}, \bibinfo{person}{Steven
  Rennie}, \bibinfo{person}{Larry~S Davis}, \bibinfo{person}{Kristen Grauman},
  {and} \bibinfo{person}{Rogerio Feris}.} \bibinfo{year}{2018}\natexlab{}.
\newblock \showarticletitle{Blockdrop: Dynamic inference paths in residual
  networks}. In \bibinfo{booktitle}{\emph{Proceedings of the IEEE conference on
  computer vision and pattern recognition}} \emph{(\bibinfo{series}{CVPR
  2018})}. \bibinfo{publisher}{IEEE}, \bibinfo{address}{Salt Lake City, Utah},
  \bibinfo{pages}{8817--8826}.
\newblock
\urldef\tempurl%
\url{https://doi.org/10.1145/10.1109/CVPR.2018.00919}
\showDOI{\tempurl}


\bibitem[Xu and Parnas(1990)]%
        {scheduling_exclusion_relation}
\bibfield{author}{\bibinfo{person}{Jia Xu} {and} \bibinfo{person}{David~Lorge
  Parnas}.} \bibinfo{year}{1990}\natexlab{}.
\newblock \showarticletitle{Scheduling processes with release times, deadlines,
  precedence and exclusion relations}.
\newblock \bibinfo{journal}{\emph{IEEE Transactions on software engineering}}
  \bibinfo{volume}{16}, \bibinfo{number}{3} (\bibinfo{date}{March}
  \bibinfo{year}{1990}), \bibinfo{pages}{360--369}.
\newblock
\urldef\tempurl%
\url{https://doi.org/10.1109/32.48943}
\showDOI{\tempurl}


\bibitem[Yang et~al\mbox{.}(2015)]%
        {googlenet-car}
\bibfield{author}{\bibinfo{person}{Linjie Yang}, \bibinfo{person}{Ping Luo},
  \bibinfo{person}{Chen Change~Loy}, {and} \bibinfo{person}{Xiaoou Tang}.}
  \bibinfo{year}{2015}\natexlab{}.
\newblock \showarticletitle{A large-scale car dataset for fine-grained
  categorization and verification}. In \bibinfo{booktitle}{\emph{Proceedings of
  the IEEE conference on computer vision and pattern recognition}}
  \emph{(\bibinfo{series}{CVPR 2015})}. \bibinfo{publisher}{IEEE},
  \bibinfo{address}{Boston, MA, USA}, \bibinfo{pages}{3973--3981}.
\newblock
\urldef\tempurl%
\url{https://doi.org/10.1109/CVPR.2015.7299023}
\showDOI{\tempurl}


\bibitem[You et~al\mbox{.}(2022)]%
        {you2022eyecod}
\bibfield{author}{\bibinfo{person}{Haoran You}, \bibinfo{person}{Cheng Wan},
  \bibinfo{person}{Yang Zhao}, \bibinfo{person}{Zhongzhi Yu},
  \bibinfo{person}{Yonggan Fu}, \bibinfo{person}{Jiayi Yuan},
  \bibinfo{person}{Shang Wu}, \bibinfo{person}{Shunyao Zhang},
  \bibinfo{person}{Yongan Zhang}, \bibinfo{person}{Chaojian Li},
  {et~al\mbox{.}}} \bibinfo{year}{2022}\natexlab{}.
\newblock \showarticletitle{EyeCoD: eye tracking system acceleration via
  flatcam-based algorithm \& accelerator co-design}. In
  \bibinfo{booktitle}{\emph{Proceedings of the 49th Annual International
  Symposium on Computer Architecture}} \emph{(\bibinfo{series}{ISCA 2022})}.
  \bibinfo{publisher}{ACM}, \bibinfo{address}{New York, NY, USA},
  \bibinfo{pages}{610--622}.
\newblock
\urldef\tempurl%
\url{https://doi.org/10.1145/3470496.3527443}
\showDOI{\tempurl}


\bibitem[Yuan et~al\mbox{.}(2016a)]%
        {hybrid_cloud_profit_maximization}
\bibfield{author}{\bibinfo{person}{Haitao Yuan}, \bibinfo{person}{Jing Bi},
  \bibinfo{person}{Wei Tan}, {and} \bibinfo{person}{Bo~Hu Li}.}
  \bibinfo{year}{2016}\natexlab{a}.
\newblock \showarticletitle{Temporal task scheduling with constrained service
  delay for profit maximization in hybrid clouds}.
\newblock \bibinfo{journal}{\emph{IEEE Transactions on Automation Science and
  Engineering}} \bibinfo{volume}{14}, \bibinfo{number}{1} (\bibinfo{date}{Feb.}
  \bibinfo{year}{2016}), \bibinfo{pages}{337--348}.
\newblock
\urldef\tempurl%
\url{https://doi.org/10.1109/TASE.2016.2526781}
\showDOI{\tempurl}


\bibitem[Yuan et~al\mbox{.}(2016b)]%
        {TTSA}
\bibfield{author}{\bibinfo{person}{Haitao Yuan}, \bibinfo{person}{Jing Bi},
  \bibinfo{person}{Wei Tan}, \bibinfo{person}{MengChu Zhou},
  \bibinfo{person}{Bo~Hu Li}, {and} \bibinfo{person}{Jianqiang Li}.}
  \bibinfo{year}{2016}\natexlab{b}.
\newblock \showarticletitle{TTSA: An effective scheduling approach for delay
  bounded tasks in hybrid clouds}.
\newblock \bibinfo{journal}{\emph{IEEE transactions on cybernetics}}
  \bibinfo{volume}{47}, \bibinfo{number}{11} (\bibinfo{date}{July}
  \bibinfo{year}{2016}), \bibinfo{pages}{3658--3668}.
\newblock
\urldef\tempurl%
\url{https://doi.org/10.1109/TCYB.2016.2574766}
\showDOI{\tempurl}


\bibitem[Zhang et~al\mbox{.}(2014)]%
        {harmony}
\bibfield{author}{\bibinfo{person}{Qi Zhang}, \bibinfo{person}{Mohamed~Faten
  Zhani}, \bibinfo{person}{Raouf Boutaba}, {and} \bibinfo{person}{Joseph~L
  Hellerstein}.} \bibinfo{year}{2014}\natexlab{}.
\newblock \showarticletitle{Dynamic heterogeneity-aware resource provisioning
  in the cloud}.
\newblock \bibinfo{journal}{\emph{IEEE transactions on cloud computing}}
  \bibinfo{volume}{2}, \bibinfo{number}{1} (\bibinfo{year}{2014}),
  \bibinfo{pages}{14--28}.
\newblock
\urldef\tempurl%
\url{https://doi.org/10.1109/TCC.2014.2306427}
\showDOI{\tempurl}


\end{thebibliography}

\end{document}